\documentclass[useAMS,usenatbib]{mn2e} 

\usepackage{enumitem}
\usepackage{float}
\usepackage{epsfig}
\usepackage{multirow}
\usepackage{epstopdf}
\usepackage{verbatim}
\usepackage{nicefrac}
\usepackage{url}
\usepackage{amsmath}
\usepackage{subfig}
\usepackage{aas_macros}
\usepackage{bm}
\usepackage[table,usenames,dvipsnames]{xcolor}
\usepackage{amsmath}
\usepackage{amssymb}
\usepackage{graphicx}
\usepackage{booktabs}

\makeatletter
\newcommand*{\rom}[1]{\expandafter\@slowromancap\romannumeral #1@}
\makeatother

\providecommand{\Omk}{\Omega_K}
\providecommand{\Oml}{\Omega_{\Lambda}}
\providecommand{\Omm}{\Omega_{\mathrm{m}}}
\providecommand{\Omb}{\Omega_{\mathrm{b}}}
\providecommand{\Omc}{\Omega_{\mathrm{c}}}
\providecommand{\omb}{\omega_{\mathrm{b}}}
\providecommand{\omc}{\omega_{\mathrm{c}}}

\providecommand{\rstar}{r_{*}}


\providecommand{\CAMB}{{\tt camb}}
\providecommand{\COSMOMC}{{\tt CosmoMC}}

\newcommand{\be}{\begin{equation}}
\newcommand{\ee}{\end{equation}}
\newcommand{\bs}{\begin{split}} 
\newcommand{\bea}{\begin{eqnarray}}
\newcommand{\eea}{\end{eqnarray}}

\newcommand{\gl}{G_{\rm light}} 
\newcommand{\gm}{G_{\rm matter}} 
\newcommand{\As}{A_{\rm s}}

\newcommand{\ns}{n_{\rm s}}
\newcommand{\lcdm}{$\Lambda$CDM}

\newcommand{\neff}{N_{\rm eff}}
\newcommand{\mnu}{\sum m_\nu}

\newcommand{\rUnit}{$h^{-1}$Mpc}
\newcommand{\vUnit}{$h^{-3}$Mpc$^3$}
\newcommand{\kUnit}{$h$ Mpc$^{-1}$}
\newcommand{\kUnitnoh}{Mpc$^{-1}$}
\newcommand{\Mpc}{{\rm Mpc}}

\def\beq{\begin{equation}} \def\eeq{\end{equation}}

\newbox\tablebox    \newdimen\tablewidth
\def\leaderfil{\leaders\hbox to 5pt{\hss.\hss}\hfil}

\def\endPlancktable{\tablewidth=\columnwidth 
    $$\hss\copy\tablebox\hss$$
    \vskip-\lastskip\vskip -2pt}
\def\endPlancktablewide{\tablewidth=\textwidth 
    $$\hss\copy\tablebox\hss$$
    \vskip-\lastskip\vskip -2pt}
\def\tablenote#1 #2\par{\begingroup \parindent=0.8em
    \abovedisplayshortskip=0pt\belowdisplayshortskip=0pt
    \noindent
    $$\hss\vbox{\hsize\tablewidth \hangindent=\parindent \hangafter=1 \noindent
    \hbox to \parindent{$^#1$\hss}\strut#2\strut\par}\hss$$
    \endgroup}
\def\doubleline{\vskip 3pt\hrule \vskip 1.5pt \hrule \vskip 5pt}

\usepackage{listliketab}

\author[~Johnson et al.]
{Andrew Johnson$^{1,2,}$\thanks{email: {asjohnson@swin.edu.au}},
Chris Blake$^1$, Jason Dossett$^3$, Jun Koda$^3$, \newauthor David Parkinson$^4$, Shahab Joudaki$^1$
 \\ \\
$^1$Centre for Astrophysics \& Supercomputing, Swinburne University of Technology, P.O. Box 218, Hawthorn, VIC 3122, Australia.\\ 
$^2$ARC Centre of Excellence for All-sky Astrophysics (CAASTRO)\\
$^3$INAF -– Osservatorio Astronomico di Brera, via Emilio Bianchi $46$, I--$23807$ Merate, Italy\\
$^4$School of Mathematics and Physics, University of Queensland, Brisbane, QLD 4072, Australia}

\date{\today}

\pubyear{2015}

\title[Searching for Modified Gravity]{Searching for Modified Gravity: Scale and Redshift Dependent Constraints from Galaxy Peculiar Velocities}

\begin{document}
\maketitle

\begin{abstract}
We present measurements of both scale- and time-dependent deviations from the standard gravitational field equations.
These late-time modifications are introduced separately for relativistic and non-relativistic particles, by way of the parameters 
$G_{\rm matter}(k,z)$ and $G_{\rm light}(k,z)$ using two bins in 
both scale and time, with transition wavenumber $0.01$ \kUnitnoh and redshift 1.
We emphasize the use of two dynamical probes to constrain this set of parameters, galaxy power 
spectrum multipoles and the direct peculiar velocity power spectrum, which probe fluctuations on different scales.
The multipole measurements are derived from the WiggleZ and BOSS Data Release 11 CMASS galaxy redshift surveys and the velocity 
power spectrum is measured from the velocity sub-sample of the 6-degree Field Galaxy Survey. 
We combine with additional cosmological probes including baryon acoustic oscillations, Type Ia SNe,
the cosmic microwave background (CMB), lensing of the CMB, and the temperature--galaxy cross-correlation. Using a Markov Chain Monte Carlo likelihood analysis,
we find the inferred best-fit parameter values of $G_{\rm matter}(k,z)$ and $G_{\rm light}(k,z)$
to be consistent with the standard model at the $95\%$ confidence level.
Furthermore, accounting for the Alcock-Paczynski effect, we perform joint fits for the 
expansion history and growth index gamma; we 
measure $\gamma =  0.665 \pm 0.0669$ ($68\%$ C.L) for a fixed expansion history, and 
$\gamma = 0.73^{+0.08}_{-0.10}$ ($68\%$ C.L) when the expansion history is allowed to deviate from 
$\Lambda$CDM. With a fixed expansion history the inferred value is consistent with GR at the $95\%$ C.L; 
alternatively, a $2\sigma$ tension is observed 
when the expansion history is not fixed, this tension is worsened by the combination of growth and SNe data.
\end{abstract}

\section{Introduction}\label{Introduction}

The observation of an accelerating cosmic expansion rate has likely
provided an essential clue for advancing our theories of gravitation and 
particle physics \citep{2001sddm.symp...27W}. Interpreting and 
understanding this feature of our Universe will require both observational 
and theoretical advancement. Observationally it is critical that we both 
scrutinise the standard vacuum energy interpretation and thoroughly search for 
unexpected features resulting from exotic physics.
Such features may exist hidden within the clustering patterns of galaxies, the coherent distortion 
of distant light rays, and the local motion of galaxies; searching for these features 
is the goal we pursue herein.      

Either outcome will facilitate progress: failure to detect 
unexpected features, confirming a truly constant vacuum energy, will 
give credence to anthropic arguments formulated within String Theory \citep{2003dmci.confE..26S}.
New observational signatures should then be targeted \citep[e.g.,][]{2013arXiv1309.4060B}. 
Alternatively, an observed deviation from a cosmological constant would indicate
a new dynamical dark energy component or a modification to Einstein's field equations
\citep{Clifton:2011nx,2006IJMPD..15.1753C}. Independent of observational progress, historical 
trends in science may offer an independent tool to predict the fruitfulness of 
each interpretation \citep{2014A&amp;G....55c3.12L}.

The possibility of new physics explaining the accelerating expansion 
has inspired an impressive range of alternative models. 
As such, a detected deviation from the standard model will 
not present a clear direction forwards, that is, interpreting such a 
deviation will be problematic. 
One potential solution, which we adopt, is to analyse observations
within a phenomenological model that captures the dynamics of a large range of
physical models \citep[e.g.,][]{2010PhRvD..81h3534B,2010PhRvD..81l3508D,2013MNRAS.429.2249S}.
It should be noted that not all approaches that introduce modified gravity or dark energy
invoke an artificial separation between the cosmological constant problem and 
the problem of an accelerating expansion \citep[e.g.,][]{2012JCAP...12..026C}.

To characterise the usefulness of phenomenological models we consider
their ability to describe known physical models: namely,
their commensurability \citep{Kuhn:1970}.
This property can be understood as describing the degree to which measurements 
made in one model can be applied to others.
The absence of this property implies that a measurement should 
only be interpreted in terms of the adopted model: a consistency test. Whereas
given this property one can constrain a range
of models simultaneously, alleviating the problem of having to re-analyse each model separately.

Specifically, the  model we adopt allows extensions to the standard \lcdm \, model by 
introducing general time- and scale-dependent modifications ($\gl$ and $\gm$) to 
General Relativity \citep{2010PhRvD..81l3508D}: these parameters
vary the relationship between the metric and 
density perturbations (i.e, they act as effective gravitational coupling).
In this case, the equivalence between the spatial and temporal metric perturbations 
is not imposed. The commensurability of our 
model to others can then be shown by proving 
that $\gl$ and $\gm$ capture all the new physics 
in specific modified gravity scenarios.

For example, \cite{2011PhRvD..83d3515D} show that by introducing parameters equivalent to
$\gl$ and $\gm$ one can provide an effective description of the entire Horndeski class of 
models. Importantly, the Horndeski class of models contains
the majority of the viable Dark Energy (DE) and modified gravity (MG) models 
\citep{2013PhRvD..87j4015S,2011PhRvD..84f4039D}. 
An often disregarded caveat is that the 
mappings between these gravitational parameters and MG and DE theories are only derived at linear order.
Therefore, until proved otherwise, the ability of the phenomenological models 
to describe physical models is
lost when using observations influenced by non-linear physics. 
To avoid this reduction in applicability we 
will focus on observations in the linear regime. We note this point 
has been emphasized elsewhere by, for example, \citet{2007APh....28..481L} and \citet{2014MNRAS.439.3504S}.

In pursuit of deviations from the standard model we use a range of 
cosmological observations.
In particular, two dynamical probes will be emphasised: the galaxy multipole power spectrum
and velocity power spectrum \citep[for example,][]{2014MNRAS.443.1065B,2014MNRAS.444.3926J}.
Hitherto, in the context of phenomenological models with scale-dependence, neither probe has been 
analysed self-consistently.
In addition we utilize the following cosmological probes: baryon acoustic oscillations, Type Ia SNe,
the cosmic microwave background (CMB), lensing of the CMB,
and temperature-galaxy cross-correlation (this correlation is caused by the Integrated Sachs--Wolf effect).

We adopt this combination of probes, direct peculiar velocities (PVs) and 
redshift-space distortions (RSDs),
to maximise our sensitivity to a range of length scales. This range is extended
as the sensitivity of both measurements 
is relatively localised at different length scales: redshift-space 
distortions at small scales, and peculiar velocity measurement 
at large scales \citep{Dodelson-Cosmology-2003}.
The benefit is an increased sensitivity to scale-dependent modifications. 
The properties of, and physical motivations for, scale-dependent modifications to 
GR are discussed by \citet{2013PhRvD..87j4015S}, and \citet{2014arXiv1409.8284B}.

In Section \ref{Sec1:Modelling deviations} we summarise  
the adopted phenomenological models and further motivate their use.
Then in Section \ref{Sec2:datasets and meth} we outline the primary datasets used 
along with the methodology we use to analyse them. Section \ref{Sec3:sec_datasets} then
presents the secondary datasets we employ.
The results and interpretations of the MCMC analysis are presented in Section \ref{Sec3:MCMC analysis}, and
the conclusions are outlined in Section \ref{sec:conclusions}.

\section{Modified Growth \& Evolution}\label{Sec1:Modelling deviations}
\subsection{Introduction}
Working within the conformal Newtonian gauge, perturbations to the Robertson-Walker metric can be 
characterised by two scalar potentials. One scalar potential describes a temporal perturbation to the metric $\psi$,
the other a spatial perturbation, $\phi$. The line element in this case is given by
\be
ds^{2} = a^2[-(1 + 2\psi)d\tau^{2} + (1 - 2\phi)d\vec{x}^2] \;,
\label{eqn:metric}
\ee
where $a$ is the scale factor, $\tau$ is the conformal time --
related to the proper time of co-moving observers by $\tau = \int dt/a(t)$ -- and 
$\vec{x}$ the spatial coordinate. A non-relativistic 
fluid within this space-time is characterised in terms of a
velocity divergence $\theta(\vec{x},\tau)$ and a density perturbation $\delta\rho(\vec{x},\tau)$.
The cosmic evolution of this fluid is then determined by its coupling to the metric potentials.

We concentrate on modifying two of the four gravitational field equations, by requiring
energy-momentum conservation ($\nabla^{\mu}T_{\mu \nu} = 0$), or equivalently, by requiring the
contracted Bianchi identity to hold, i.e., $\nabla^{u}G_{\mu \nu} = 0$.
Enforcing either constraint one finds the relativistic continuity and Euler equations in Fourier space:
\begin{eqnarray}
\dot{\delta}_m &=&- \theta_m + 3 \dot{\phi},  \label{SE1} \\
\dot{\theta}_m &=& - \mathcal{H} \theta_m + k^2 \psi \, .   \label{SE2}
\end{eqnarray}
where $\delta_m \equiv \delta \rho_{m} / {\bar \rho_m}$ and $\mathcal{H} \equiv \dot{a}/a = (da/d\tau)/a$, and 
${\bar \rho_m}$ is 
the background matter density. 
This system of four variables can then be closed by specifying the gravitational field equations;
in particular, by defining 
the relationship between the two metric potentials, and the coupling 
between the metric potentials and the matter over-density.
In GR these relationships are given by
\begin{eqnarray} 
\nabla^2 \psi&=&4\pi G_N a^2 {\bar \rho_m} \Delta_m  \\ 
\phi &=& \psi \, ,
\label{eqn:GRfield}
\end{eqnarray}
where $G_{\rm N}$ is Newton's gravitational constant, and the equations
are defined in terms of the comoving-gauge density perturbation $\Delta_{m} = \delta_m  + (3\mathcal{H}/k^{2})\theta_{m}$.

\subsection{$\gl(k,z)$ and $\gm(k,z)$}

We now introduce two dimensionless free parameters $\gl$ and $\gm$ that 
we use to model deviations to the field equations. Our model is now specified as \citep{2013JCAP...02..007D}
\begin{eqnarray} 
\nabla^2 \psi&=&4\pi G_N a^2 {\bar \rho_m} \Delta_m \times G_{\rm matter} \\ 
\nabla^2(\phi+\psi)&=&8\pi G_N a^2 {\bar \rho_m} \Delta_m \times G_{\rm light} \ . 
\label{eqn:GR modifications}
\end{eqnarray}
The first equation governs the motion of non-relativistic particles, while the second
controls the propagation of light along null geodesics. As a result, $G_{\rm matter}$
can be measured using RSDs and direct PVs, and $G_{\rm light}$ can be measured using weak lensing.
Because of this distinction the two parameters are significantly less correlated than models involving a `slip' relation
\citep[e.g.,][]{2010PhRvD..81h3534B}.
Note that the variables $\{\Sigma,\mu\}$ in \cite{2013MNRAS.429.2249S} and \cite{2012PhRvD..85l3546Z} are
equivalent to $\{\gl,\gm\}$. There is also a trivial re-mapping to the 
$\{Q,R\}$ parameters used by \citet{2010PhRvD..81h3534B}, through $\gm=QR$, $\gl=Q(1 + R)/2$.

To ensure our model can test for a variety of deviations from GR we allow for 
both scale- and redshift-dependence: that is, 
$\gl = \gl(z,k)$ and $\gm = \gm(z,k)$.
To specify these parameters we use a high vs. low-redshift,
large vs. small scale binning approach introduced by \citet{2010PhRvD..82j3523D}. 
Note, however, that very general functional forms for these parameters 
(including scale-dependent terms) have been developed \citep{2013PhRvD..87j4015S,2014arXiv1409.8284B}. 
We leave such investigations to future work.

Our adopted model introduces $8$ free parameters and requires one to specify a redshift and wavenumber 
transition scale, $z_t$ and $k_t$. We set $z_t = 1$ and $k_{c} = 0.01$ \kUnitnoh; therefore, we have
two redshift bins (viz., $0<z<1$ and $1<z<2$) and two wavenumber bins
($10^{-4}{\rm Mpc}^{-1}<k<10^{-2}{\rm Mpc}^{-1}$ and 
$0.01\,{\rm Mpc}^{-1}<k<0.1\,{\rm Mpc}^{-1}$), while for $z>2$ and $k < 10^{-4}{\rm Mpc}^{-1}$  GR is restored.  
The transition between 
bins is implemented using an $\arctan$ function of width $\Delta z = 0.05$ and $\Delta k = 0.001$. 

For our first model we choose to leave the cosmic expansion unmodified at the $\Lambda$CDM prediction, 
and concentrate on the growth of structure.
Henceforth, we will refer to this model as {model} {\color{red}\rom{1}}.
To calculate the relevant observables (to be discussed in the next section) we use
$\CAMB$ and $\COSMOMC$. The modified field equations (Eq \ref{eqn:GR modifications}) are incorporated into $\CAMB$
using the publicly available code {\tt ISITGR} \citep{2011PhRvD..84l3001D}, and
the exact equations implemented in $\CAMB$ are given by \citet{2011PhRvD..84l3001D}. Note the only 
significant difference between the equations employed in $\CAMB$ and Eq (\ref{eqn:GR modifications}) is that the latter
are written within the synchronous gauge \citep{1995ApJ...455....7M}. 

A few technical comments on the model are unavoidable:
Firstly, super-horizon curvature perturbations need to be conserved independent of the form of field equations \citep{2008PhRvD..78b4015B}. 
This condition was shown to be satisfied for this model by \citet{2010PhRvD..81j4023P}. 
Additionally, it is natural to include a smoothness theory prior on these parameters, however,
given the large distance between the centre of our bins we choose not to include such a prior \citep{2013PhRvD..87j4015S}. 
With more accurate data, and hence a larger number of bins, this argument will no longer be valid.
Finally, the accuracy of any mapping from our model to physical models (i.e., those derived from an action) relies on the validity of 
the quasi-static approximation (QSA). 
Following the arguments presented in \citet{2013PhRvD..87j4015S} it is reasonable to 
include a theoretical prior to ignore such deviations.

\subsection{Varying Growth and Expansion: $\{\gamma,w_{0},w_{a}\}$}

As more freedom is introduced to model deviations from GR the precision of the inferred parameters 
degrades. We must decide then which features of the standard model to preserve; for example, to what 
extent does the expansion history dictate the growth history.
This presents a balancing problem with no clear solution. 
To partially circumvent this issue we adopt a second model (which we label model {\color{red}\rom{2}}). 
In contrast to our first model, this model includes only minimal extensions to the standard model. 
As a result there are fewer free parameters and more precise tests are possible (although we nonetheless introduce deviations to both the expansion and growth history).

This minimal extension to the standard model using the parameters $\{w_{0},w_{a},\gamma\}$ has been advocated 
by \citet{2007APh....28..481L,Linder:2005kl}, and \cite{2010PhRvD..81d3512S}, and applications 
have been presented, for example, by \citet{2007PhRvD..75b3519H}.
To expand on this, we introduce deviations to the expansion history through
a time-dependent equation of state $w(z)$, which is expressed in terms of two free parameters:
$w_{0} = w(a = 0)$ and $w_{a} = -(d w/d a)\big|_{a = 1}$,
as a function of the redshift $w(z) = w_{0} + w_{a}z/(1 + z)$.
Note the expansion history is still governed by the Friedman equation, there is simply 
more freedom in the properties of the dark energy component.
We introduce deviations in the growth history 
by parameterizing the growth rate as  $f(z) \equiv \Omega_m(z)^{\gamma}$,
where $\gamma$ is the growth index; within GR one expects $\gamma \sim 0.55$. The growth rate is defined by
$f(a) \equiv d \ln D(a) /d \ln a $, and $ D(a) \equiv \delta(a)/ \delta(a=1)$.

\section{Primary Datasets: Methodology}\label{Sec2:datasets and meth}

Below we will outline the measurements we use in Sec.~\ref{Sec3:MCMC analysis}, in addition to 
the tools we use to analyze them. A general summary is provided in Table \ref{tab:datasets} where
the datasets, the measured quantities, and the fitting ranges adopted are specified. The focus will be on 
introducing extensions to the public MCMC code $\COSMOMC$ \citep{2002PhRvD..66j3511L} and 
$\CAMB$ \citep{Lewis:2000nx} to update the range of datasets one can analyze.

\begin{table*}
\begin{center}
\caption{Summary of the datasets used in this analysis. 
Given model {\color{red}\rom{1}} includes scale-dependent terms, we
divide our measurements into three separate groups: those used to constrain model {\color{red}\rom{1}} \& {\color{red}\rom{2}}, 
only model {\color{red}\rom{1}}, and only model {\color{red}\rom{2}}. This division is indicated by the horizontal lines, 
and follows the order in which the categories were introduced.}
\begingroup 
\newdimen\tblskip \tblskip=5pt
\label{tab:datasets}
\vskip -5mm
\footnotesize 
\setbox\tablebox=\vbox{ %
\newdimen\digitwidth 
\setbox0=\hbox{\rm 0}
\digitwidth=\wd0
\catcode`*=\active
\def*{\kern\digitwidth}
\newdimen\signwidth
\setbox0=\hbox{+}
\signwidth=\wd0
\catcode`!=\active
\def!{\kern\signwidth}
\halign{\hbox to 2.7cm{#\leaderfil}\tabskip=0.7cm & \hfil#\hfil\tabskip=0.5cm &  \hfil#\hfil\tabskip=0.2cm &#\hfil\tabskip=0pt\cr
\noalign{\doubleline}
\omit\hfil Cosmological Probe \hfil&\omit\hfil Dataset \hfil&\omit\hfil Measured quantity \hfil&\omit\hfil Reference \hfil\cr
\noalign{\vskip 3pt\hrule\vskip 3pt}
CMB temperature\dots & {Planck} & $C^{T\,T}_{l}$ & \citet{Collaboration:2013qf}\cr
CMB polarization\dots & {WMAP-}$9$ & $C^{E\,E}_{l}$ &\citet{2013ApJS..208...20B}\cr
CMB-Lensing\dots            &{Planck}& $C^{\phi\,\phi}_{l}$& \citet{2013arXiv1303.5077P} \cr
BAOs & $6$dFGS & $r_s/D_{V}(z)$ & \citet{2011MNRAS.416.3017B} \cr
 &BOSS DR$11$ LOWZ & $D_{\rm V}(r_s^{{\rm fid}}/r_{s})$ & \citet{2013arXiv1303.4666A} \cr
 &BOSS DR$11$ QSA-Ly$\alpha$ & $H(z)r_{s},D_{A}/r_{s}$ &  \citet{2014JCAP...05..027F} \cr
 &BOSS DR$11$ Ly$\alpha$ & $H(z)r_{s},D_{A}/r_{s}$ & \citet{2014arXiv1404.1801D} \cr
Type Ia Supernovae \dots & {SNLS} & $\mu(z)$  &\citet{2011ApJS..192....1C} \cr
\noalign{\vskip 3pt\hrule\vskip 3pt}
Dataset extension {\color{red}\rom{1}}            &   & &\cr
ISW-density cross\dots       &{WMAP3} & $C^{g\,T}_{l}$       & \citet{2008PhRvD..78d3519H} \cr
Velocity Power Spectrum & $6$dFGSv & ${\mathcal P}_{vv}(k)$   & \citet{2014MNRAS.444.3926J} \cr
BAO (reconstructed)$^{\rm a}$ & WiggleZ     & $D_{\rm V}(r_s^{{\rm fid}}/r_{s})$ & \citet{2014MNRAS.441.3524K} \cr
& DR$11$ CMASS & $D_{\rm A}(z)(r_{\rm s}^{\rm fid}/r_{\rm s}) , H(z)(r_{\rm s}/r_{\rm s}^{\rm fid}) $  & \citet{2014MNRAS.441...24A} \cr
Power Spectrum Multipoles & DR$11$ CMASS & $P_{0}(k),P_{2}(k)$      & \citet{2014MNRAS.443.1065B} \cr
& WiggleZ $(z_{\rm eff}=0.44)$ & $P_{0}(k),P_{2}(k),P_{4}(k)$  & \citet{2011MNRAS.415.2876B}$^{b}$ \cr
& WiggleZ $(z_{\rm eff}=0.73)$ & $P_{0}(k),P_{2}(k),P_{4}(k)$ & \citet{2011MNRAS.415.2876B}$^{b}$ \cr
\noalign{\vskip 3pt\hrule\vskip 3pt}
Dataset extension \color{red}\rom{2}            &   & &\cr
RSDs & 6dFGS    & $f\sigma_{8}(z)$   & \citet{Beutler:2012fk} \cr
RSD-BAO-AP & WiggleZ      & $A(z),F_{\rm AP}(z),f\sigma_{8}(z)$        & \citet{2012MNRAS.425..405B} \cr 
RSD-BAO-AP & BOSS CMASS   & $D_{\rm v}/r_{\rm s}(z),F_{\rm AP}(z),f\sigma_{8}(z)$ & \citet{2013arXiv1312.4611B} \cr
\noalign{\vskip 3pt\hrule\vskip 3pt}}}
\endPlancktablewide 
\tablenote \textit{a} Both the reconstructed BAO measurements  (CMASS and WiggleZ) have been calculated by marginalising over 
the general shape of the correlation function. Marginalising over the shape decorrelates the BAO measurement with the power spectrum multipole measurement, 
allowing one to fit for both measurements simultaneously.\par
\tablenote \textit{b} Note, however, these measurement have been updated in this work using an improved methodology. \par
\endgroup
\end{center}
\end{table*}

\subsection{Velocity Power Spectrum}

The radial PVs of galaxies in the local universe
induce a fluctuation in the apparent magnitude $m$, defined as \citep{Hui:2006fk}
\begin{eqnarray}
\delta m(z) = [m(z) -\bar{m}(z)] \, .
\end{eqnarray}
The over-bar indicates that the variable is being evaluated 
within a homogeneous universe, namely, a universe with no
density gradients and therefore no peculiar velocities. Recall the apparent magnitude is defined as
\begin{eqnarray}
m = M + 5\log_{10}(D_{\rm L}(z)) + 25\,.
\label{mag}
\end{eqnarray}
Here $M$ is the absolute magnitude, and $D_{\rm L}(z)$ the luminosity distance.
The presence of large scale clustering induces fluctuations in $\delta m(z)$ from galaxy to galaxy (this is equivalent to a peculiar velocity), furthermore, 
these fluctuations are correlated for nearby galaxies \citep{Hui:2006fk,Gordon:2007dq}.
The magnitude of both effects can be
described by a covariance matrix which we 
define as $C^{\rm m}_{i j} \equiv \langle\delta m_{i}(z_{i}) \delta m_{j}(z_{j})\rangle$.
Once a model is specified this covariance matrix can be calculated as 
\begin{equation}
C^{\rm m}_{i j } = G(z_{i},z_{j})\int \frac{d k}{ 2\pi^{2}}k^{2} {\cal P}_{ v v}(k,a=1)W(k,\alpha_{i j},r_{i},r_{j})\, .
\label{eqn:vcov}        
\end{equation}
Where ${\cal P}_{ v v}(k) = {\cal P}_{ \theta \theta}(k)/k^2$ is the velocity power spectrum, and $\theta = \nabla \cdot \vec{v}$ is the velocity 
divergence, furthermore
\beq
\begin{split}
&W(k,\alpha_{i j},r_{i},r_{j})  =  1/3 \left[j_{0}(kA_{ij}) - 2j_{2}(kA_{ij})\right]\hat{r}_{i} \cdot \hat{r}_{j}  \\
& + \frac{1}{A_{ij}^2}j_{2}(kA_{ij})r_{i}r_{j}\sin^{2}(\alpha_{ij}) \,, \\ 
&G(z_{i},z_{j}) \equiv \\ 
&\left( \frac{5} {\ln{10} }\right)^{2}  \left( 1 - \frac{(1 + z_{i})^{2}}{H(z_{i})D_{\rm L}(z_{i})}\right)   \left( 1 - \frac{(1 + z_{j})^{2}}{H(z_{j})D_{\rm L}(z_{j})}\right) \, , \nonumber
\end{split}
\eeq
where $\alpha_{ij} = \cos^{-1}(\hat{r}_{i} \cdot \hat{r}_{j})$, $A_{ij} \equiv | {\bf r}_{i} - {\bf r}_{j}|$
and ${\bf r}_{i}$ is the position vector of the i$^{\text{th}}$ galaxy. This analytic solution 
for the window function was presented by \citet{Ma:2010bh}. For further details on this 
calculation we refer the reader to \citet{2014MNRAS.444.3926J}.

We perform a 
full likelihood calculation using the 6dFGSv peculiar velocity sample \citep{2014MNRAS.445.2677S}.
To calculate the covariance matrix in Eq. (\ref{eqn:vcov}) we integrate over the wavenumber range 
$k=0.0005-0.15$\kUnit. Given the dominance of large-scale information in peculiar velocity measurements,
we neglect velocity bias in this calculation.

In order to minimise the influence of poorly understood non-linear effects
a non-linear velocity dispersion component $\sigma_{\rm PV}$ is introduced into the 
diagonal elements of the covariance matrix \citep{Silberman:2001ve}. This nuisance parameter 
is marginalised over in the analysis. The covariance matrix is thereby updated:
\begin{eqnarray}
\Sigma_{i j} \equiv C^{\rm m}_{i j } + \sigma^{2}_{\rm PV}\delta_{i j} \; .
\label{Sig} 
\end{eqnarray}
One can now define the posterior distribution as
z\beq
P(\Sigma | \delta {\bf m}) = |2\pi \Sigma |^{-1/2}\exp{\left(-\frac{1}{2}{\bf \delta m}^{\text{T}} \Sigma^{-1}{\bf \delta m} \right)},
\label{like}
\eeq
where ${\bf \delta m}$ is a vector of the observed apparent magnitude fluctuations. 
Note the dependence on the cosmological model is introduced through the covariance matrix. 

The model velocity power spectrum is generated using a transfer function. This can be defined starting from 
the peculiar velocity in the synchronous gauge $v_{\rm p}^{(\rm s)}$ \citep[cf.,][]{1995ApJ...455....7M}\footnote{
Our starting point is set by variables used within $\CAMB$.}. 
As this gauge is defined in the dark matter rest frame, i.e., there are no temporal $g_{0 0}$ perturbations,
a gauge transformation is necessary. Using 
the convention of \citet{1995ApJ...455....7M} we define $h$ and $\eta$ as the metric perturbation in the synchronous gauge.
Now by moving into the Newtonian gauge one finds the appropriate transfer function:
\beq
T_{v}(k) =  \frac{c}{k^{2}}\left( k \alpha + \rho_b \, v_p^{(s)}/(\rho_b + \rho_c)\right),
\eeq
where $k^{2}\alpha = \dot{h}/2 + 3\eta$.

\begin{figure*}
\centering
\includegraphics[width=15cm]{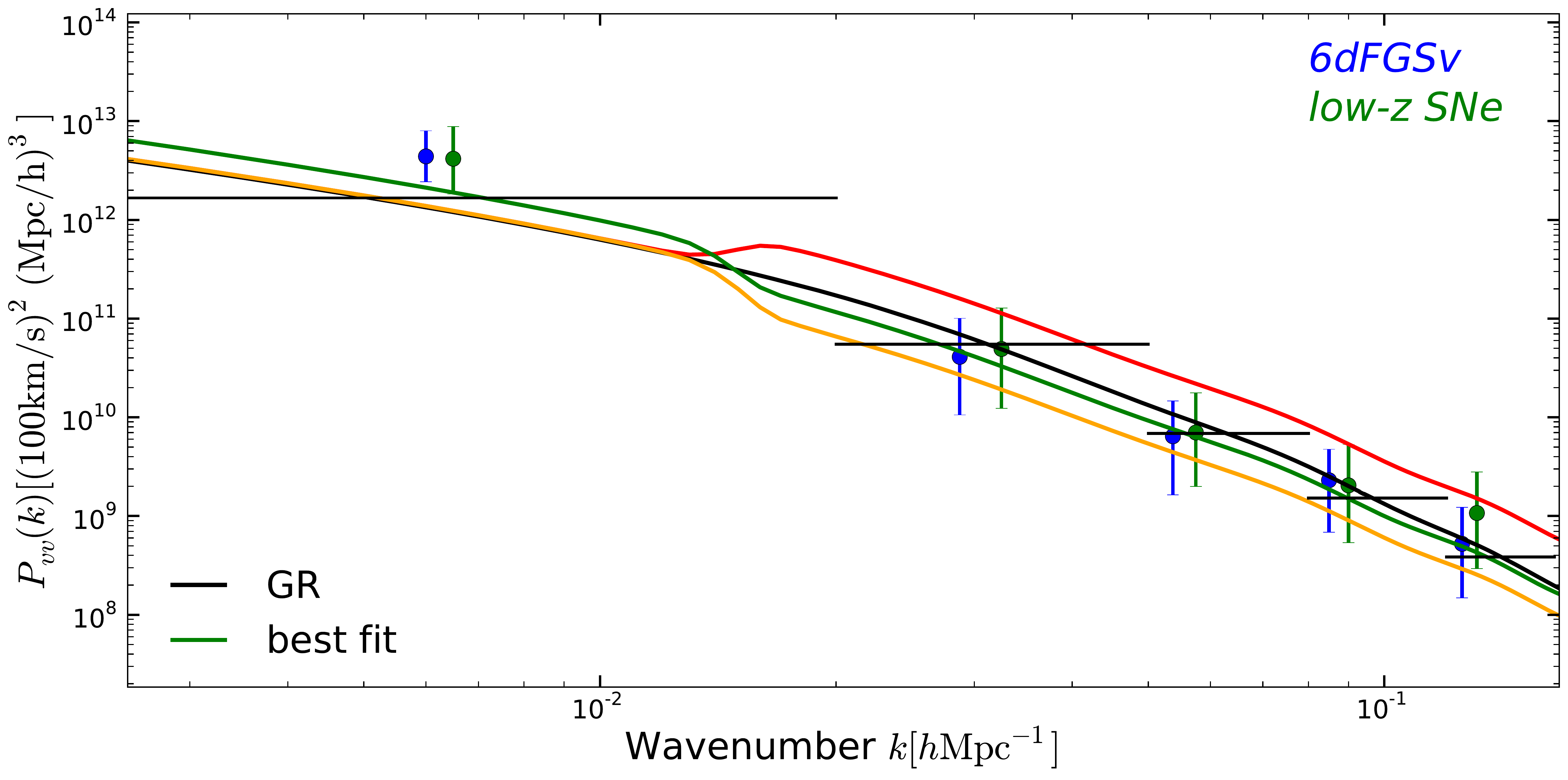}
\caption{The velocity power spectrum ${\mathcal P}_{\rm v v}(k)$ at $z=0$ for different parameter 
combinations of the adopted phenomenological model.
The black line shows the prediction assuming General Relativity, and
the orange and red lines illustrate the effect of varying the 
low-$z$ and high-$k$ bin for $\gm$. For the red line $G_{\rm matter}(z<1;k>0.01) = 1.8$ and for 
the orange line $G_{\rm matter}(z<1;k>0.01) = 0.3$: for these predictions
the standard cosmological parameters are fixed at the Planck best-fit values, and unless specified 
otherwise all non-GR parameters are set to be consistent with GR (i.e., set equal to $1$).
Moreover, the green line shows the prediction found using the best-fit 
parameter values found using set 4 (see sect. \ref{Sec3:MCMC analysis} for details). 
The best-fit values here correspond to the parameter values that maximise the likelihood.
The blue and green data points correspond to the $68\%$ confidence intervals 
for the mean power within each bin for the $6$dFGSv data and the
low-$z$ SNe data set constructed in \citet{2014MNRAS.444.3926J}. 
The thick black line indicates the the mean power predicted by GR in each k-bin, this is calculated assuming a 
Planck cosmology.
}
\label{Plot:Pvv}
\end{figure*}

In Fig. \ref{Plot:Pvv} we plot the measurements of ${\mathcal P}_{vv}(k)$ by \citet{2014MNRAS.444.3926J},
here the blue (green) points were measured using the 6dFGSv (low-$z$ SNe) sample.
For this plot the black line shows the power spectrum prediction assuming GR, while
the red and orange lines show the predictions for different values of the post-GR parameters.
For these calculations the {\it Planck} best-fit parameters are assumed.
Additionally, the green line shows the prediction when using our best-fit 
parameter values (see sect. \ref{Sec3:MCMC analysis} for details).
Note the time evolution of the density perturbation $\Delta_{m}$ is set by 
a friction term $2{\mathcal H}\Delta_{m}$ and a source term $k^{2}\psi$.
Therefore, by modifying $G_{\rm matter}$ one changes the source term to
$k^{2}\psi \sim − a^{2} G_{\rm matter}(k,z)\Delta_{m}$; hence,
with $G_{\rm matter}(k,z) > 1$ both the late-time clustering and the 
amplitude of the velocity power spectrum are enhanced.

\subsection{Power Spectrum Multipoles}

We measured the multipole power spectra of the WiggleZ Survey data using 
the direct estimation method introduced by \citet{2006PASJ...58...93Y} and 
extended by \citet{2011MNRAS.415.2876B} and \citet{2014MNRAS.443.1065B}. We 
provide a brief summary of the technique here, referring the reader to 
the above papers for a full description.

The redshift-space 2D galaxy power spectrum $P_g^s(k,\mu)$, where $\mu$ 
is the cosine of the angle of the wavevector $\vec{k}$ with respect to 
the line-of-sight, may be expressed in terms of multipole moments 
$P_\ell(k)$ using a basis of Legendre polynomials $L_\ell(\mu)$:
\begin{equation}
P_g^s(k,\mu) = \sum_{{\rm even} \, \ell} P_\ell(k) \, L_\ell(\mu) ,
\end{equation}
where
\begin{equation} 
P_\ell(k) = \frac{2\ell + 1}{2} \int_{-1}^1 d\mu \, P_g^s(k,\mu) \, 
L_\ell(\mu) .
\label{eqpkl}
\end{equation}
The power spectrum multipoles provide a form of data compression; in 
linear theory all the information is contained in the $\ell = 0,2,4$ 
terms, with the first two multipoles dominating the observed signal.

The rapid estimation technique of using Fast Fourier Transform (FFT) 
methods to measure $P_g^s(k,\mu)$ in bins of $k$ and $\mu$, where $\mu$ 
is defined with respect to a fixed axis parallel to the line-of-sight of 
the field centre, and then estimating $P_\ell(k)$ by a direct sum over 
the binned results using Equation \ref{eqpkl}, has two difficulties.  
First, for a wide-area survey the line-of-sight direction with respect 
to which $\mu$ should be measured will not be fixed.  Secondly, at low 
$k$ the sum over $\mu$ bins is problematic to evaluate due to the 
limited number of modes available in Fourier space. The \citet{2006PASJ...58...93Y}
method estimates $P_\ell(\vec{k})$ using a sum over all 
galaxies for each wavevector $\vec{k}$ on the FFT grid, allowing the 
line-of-sight vector to vary for each object and without binning in 
$\mu$.  Window function effects are included using a similar sum over 
unclustered objects.  Additive corrections are included for shot noise 
and for the discreteness of the grid.  The measurements are then binned 
by wavenumber $k = |\vec{k}|$.

\begin{figure*}
\centering
\includegraphics[width=17.5cm]{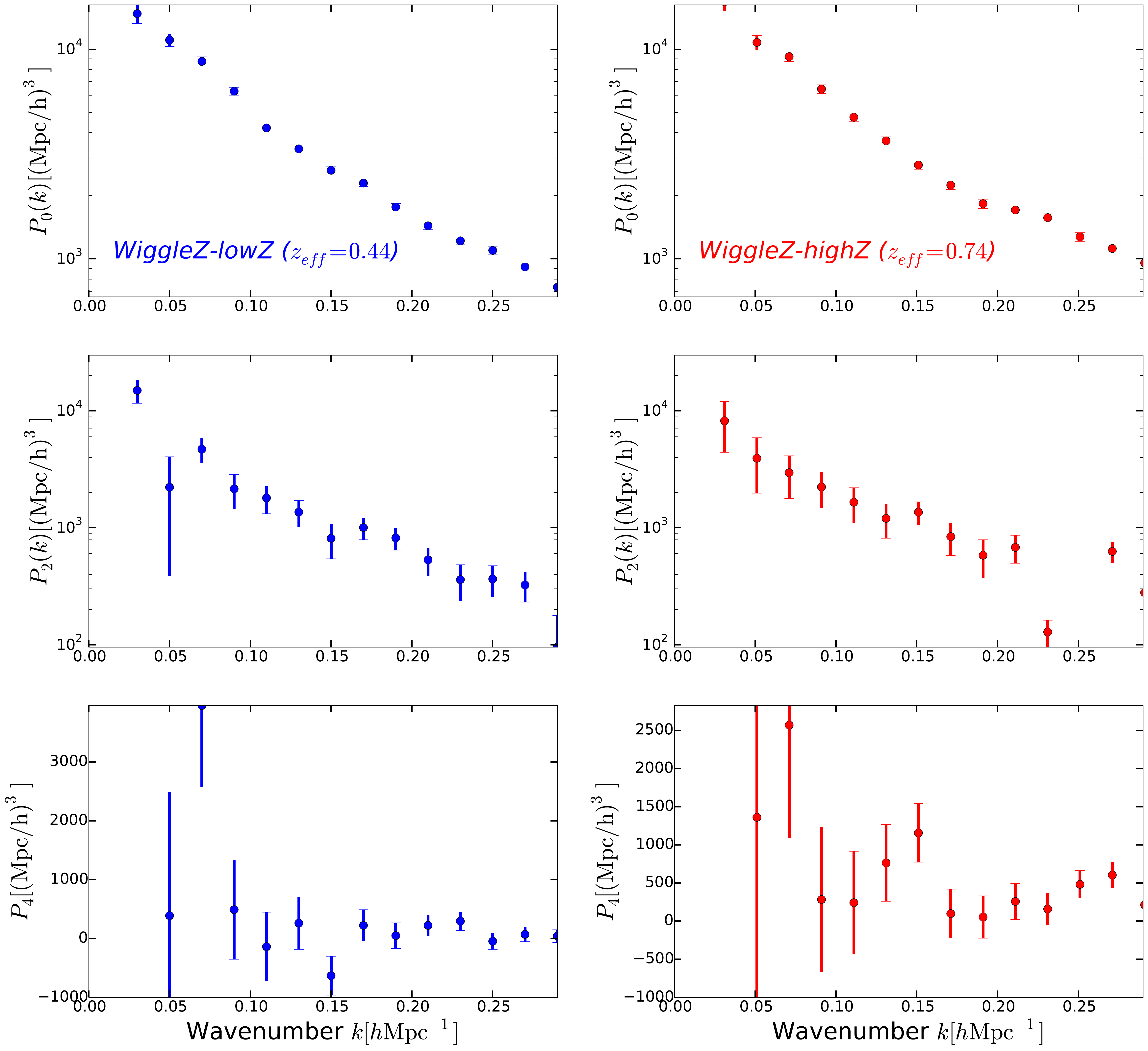}
\caption{The monopole, quadrupole and hexadecapole measurements from the WiggleZ survey, for both the high-$z$ and low-$z$ samples ($z = 0.44,0.73$, respectively). For simplicity 
we combine the results from the $6$ different survey regions; however, note this is not the format of the data we use: each survey region has a different window function and 
therefore is analysed separately.}
\label{Plot:Wigglez}
\end{figure*}

Following the analysis of the WiggleZ baryon acoustic oscillations 
\citep{Blake:2011kl}, we estimated the $\ell=0,2,4$ multipole power 
spectra in the $(9,11,15,22,1,3)$-hr survey regions in the overlapping 
redshift ranges $0.2 < z < 0.6$, $0.4 < z < 0.8$ and $0.6 < z < 1.0$.  
We measured the spectra in 14 wavenumber bins of width $\Delta k = 0.02 
\, h$ Mpc$^{-1}$ in the range $0.02 < k < 0.3 \, h$ Mpc$^{-1}$. For this 
analysis, however, we only use the non-overlapping redshift ranges that we 
label low--z and high--z. The results for the monopole and
quadrupole are given in Fig. \ref{Plot:Wigglez}.

We determined the covariance matrix of each vector $[P_0(k), P_2(k), 
P_4(k)]$ by repeating the measurements in each survey region for a 
series of 600 mock catalogues, built from N-body simulations generated 
by the method of COmoving Lagrangian Acceleration \citep[COLA;][]{2013JCAP...06..036T}.
As described by \citet{2014MNRAS.441.3524K}
we produced a halo catalogue by applying a friends-of-friends algorithm 
to the dark matter particles, and populated the haloes with mock 
galaxies using a Halo Occupation Distribution such that the projected 
clustering matched that of the WiggleZ galaxies.  The mocks were 
sub-sampled using the selection function of each region, and galaxy 
co-ordinates converted to redshift-space.

We also determined the convolution matrix for each region and redshift 
slice, which should be used to project a model multipole vector to form 
a comparison with the data given the survey window function.  For a 
wide-angle survey such as the BOSS, determination of the convolution 
involves a numerically-intensive double sum over randomly-distributed 
objects \citep{2014MNRAS.443.1065B}.  However, for the more compact WiggleZ 
Survey geometry, we found that it was acceptable (in the sense that any 
offset was far smaller than the statistical error) to use a flat-sky 
approximation, in which FFT methods were used to convolve a series of 
unit multipole vectors, generating each row of the convolution matrix in 
turn.

In addition to the WiggleZ multipole measurements, we include the monopole and 
quadrupole measurements from the BOSS-DR11 CMASS sample presented in \cite{2014MNRAS.443.1065B}; the reader
is referred to this paper for technical details on the calculation. From the CMASS sample the 
$l = 0,2$ multipole power spectrum are calculated for the wavenumber range $k=0.01-0.20$\kUnit with  
a spacing of $\Delta k=5\times 10^{-3}$\kUnit. These measurement are presented for both 
the North and South Galactic Cap regions at an effective redshift of 
$z_{\rm eff} = 0.57$. 

\begin{figure*}
\centering
\includegraphics[width=17cm]{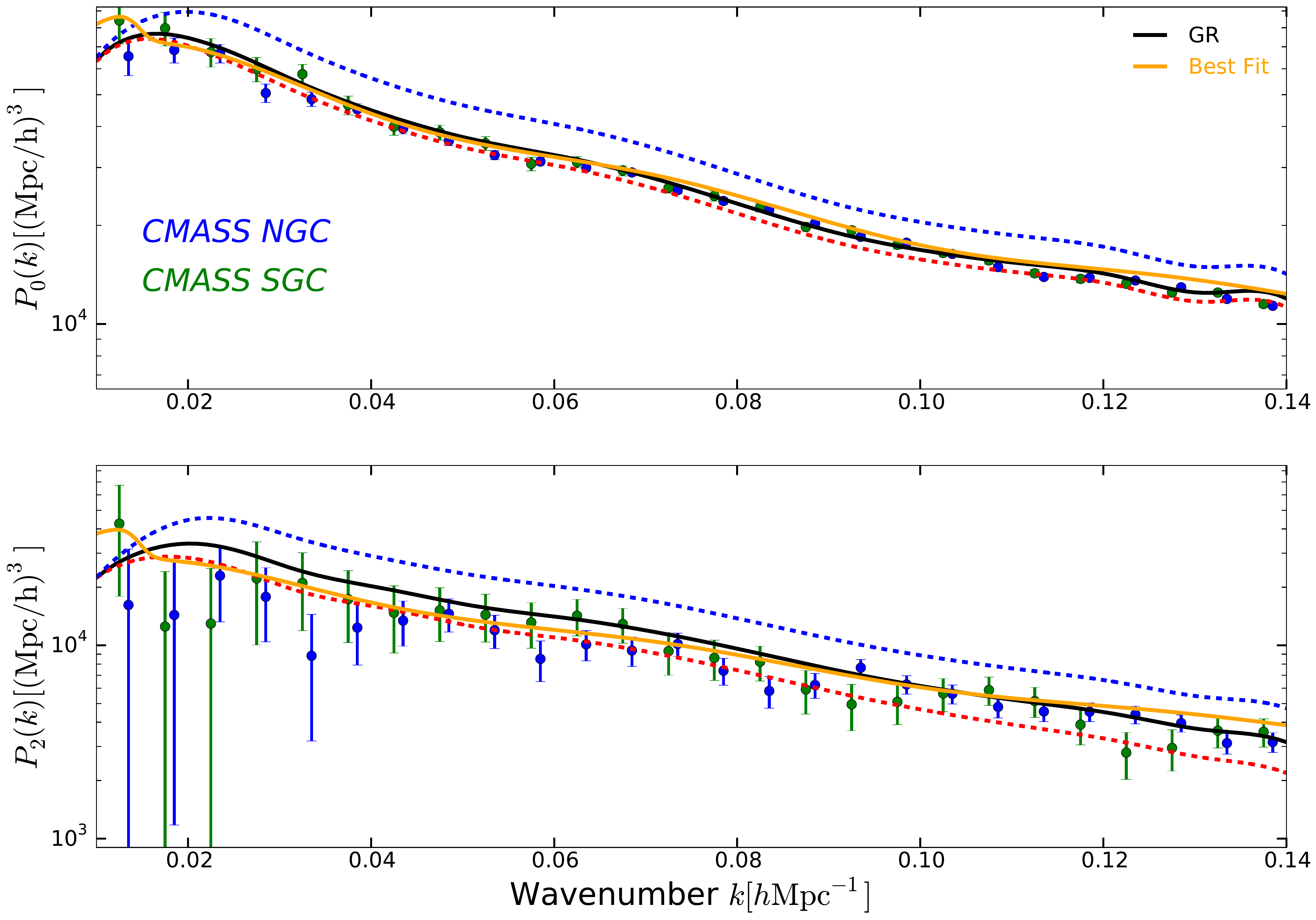}
\caption{The monopole and quadrupole power spectrum for both the BOSS-DR11 CMASS survey regions (NGC and SGC). 
The blue-dashed line shows the prediction with $G_{\rm matter}(k>0.01;z>1) = 1.8$, the red-dashed lines
$G_{\rm matter}(k>0.01;z<1) = 0.3$; for these predictions the best-fit parameters from Planck are assumed
and we set the bias to $b = 1.85$, shot noise to $N = 1800\,$\vUnit, and the velocity dispersion to $\sigma_{v}=4\,$\rUnit.
For simplicity these theory predictions have only been convolved with the NGC window function. The orange lines 
gives the prediction from the best-fit model parameters (see sect. \ref{Sec3:MCMC analysis} for details), convolved with the NGC window function.
Note, for the final analysis we only fit our model to $k_{\rm max} = 0.10$\kUnit .}
\label{Plot:CMASS}
\end{figure*}

We plot the CMASS multipole measurements in Fig. \ref{Plot:CMASS}. For this plot
the blue-dashed (red-dashed) lines show the multipole predictions when setting $G_{\rm matter}(k>0.01;z<1) = 1.8$ ($G_{\rm matter}(k>0.01;z<1) = 0.3$),
while the black lines show the prediction assuming GR.
For these predictions the best-fit parameters from Planck are assumed, in addition
we set the bias to $b = 1.85$, shot noise to $N = 1800\,$\vUnit, and the velocity dispersion to $\sigma_{v}=4\,$\rUnit.
Moreover, the orange lines 
give the prediction when using our best-fit model parameters (see sect. \ref{Sec3:MCMC analysis} for details)
Note, for simplicity the theory predictions have only been convolved with the NGC window function.

\subsubsection{Modelling the Power Spectrum Multipoles}\label{subsec:modelPk}

To model the redshift-space 2D galaxy power spectrum $P_g^s(k,\mu)$ we use
linear theory plus an empirical Gaussian damping term  \citep{1998MNRAS.296...10H}; 
the resulting model is given by
\be
P_g^s(k,\mu) = \left[P_{gg}(k) - 2 \mu^2 P_{g\theta}(k) + \mu^4 P_{\theta\theta}(k)\right]D(\mu,k) \; ,
\label{eqn:streaming model}
\ee
where $D(\mu,k) = \exp[-(k f\mu \sigma_{\rm v})^2]$. The standard
interpretation of this damping, which is clearly observed in redshift surveys, 
is the uncorrelated pairwise velocity dispersion of galaxies. 
We absorb our ignorance by treating $\sigma_{\rm v}$ as a free 
parameter to be marginalised over for each survey.

Assuming linear theory the continuity equation (eq. \ref{SE2}) can be written in Fourier space as
\begin{eqnarray}
\theta(k) &=& -f(a)\delta(k) \; .
\label{rsd cont}
\end{eqnarray}

However, we are modifying the gravitational field equations, so one needs to be 
self-consistent, given that the modifications (Eq. \ref{eqn:GR modifications}) will 
change the growth rate in a scale-dependent manner. We calculate this modified scale-dependent growth rate 
as
\begin{eqnarray}
f(k,a) = \frac{d \ln \Delta_{c}(k,a)}{d \ln a} \; .
\end{eqnarray}
This is self-consistent given $\CAMB$ contains all the relevant physics, i.e, the density and velocity variables are evolved according to 
the modified field equations. As a reminder of 
the potential scale-dependence we write the growth rate as $f(k)$. Now
assuming a local, scale-independent linear bias $(\delta_{\rm g} = b\delta)$ and no velocity bias ($\theta_{g} = \theta$)
Eq. (\ref{eqn:streaming model}) reduces to
\begin{eqnarray}
P_g^s(k,\mu) = b^2 \, (P_{\delta\delta}(k) + N )\left(1 + f(k)\mu^2/b \right)^2 D(\mu,k) \; ,
\label{eqn kaiser + damping}
\end{eqnarray}
Here we have included a shot noise component $N$, this is treated as a free parameter for the CMASS 
analysis.
To justify the previous assumptions we truncate the fit for both the WiggleZ and CMASS multipoles
to relatively large scales; to wit, we set $k_{\rm max}^{\rm CMASS}=0.10$\kUnit and $k_{\rm max}^{\rm WiggleZ}=0.15$\kUnit.
The WiggleZ measurements are used to a higher wavenumber because of 
the smaller bias of the sample ($b \sim 1$), in addition to
the larger error bars\footnote{With a lower biased tracer, for example, the effect of 
non-local halo bias is less significant \citep{2012PhRvD..85h3509C}.}.
The matter power spectrum is calculated within $\CAMB$ using only linear 
theory: we choose not to incorporate non-linear corrections via {\tt HALOFIT}. 
The use of {\tt HALOFIT} presents an issue as the corrections have 
not been shown to be valid for general modified gravity models.

In order to correctly interpret RSD measurements
one is required to consistently incorporate our ignorance of the expansion history of the universe (viz., $H(z)$),
bearing in mind that these measurements are performed assuming a fiducial cosmological model. 
As a result, in a trial cosmology, the growth rate measurements should be adapted using the covariance 
with the Alcock-Paczynski (AP) distortion.
Any discrepancy between the chosen fiducial expansion history ($\hat{D}_{\rm A}(z),\hat{H}(z)$) 
and the physical expansion history (${D}_{\rm A}(z), {H}(z)$) can be accounted 
for by scaling the true (physical) radial and tangential wavenumbers 
($k^{\rm true}_{\parallel},k^{\rm true}_{\perp}$).
The amplitude of the wavenumber scalings is determined by 
\begin{eqnarray}
\alpha_{\parallel} &=& \frac{H^{\rm fid}(z)}{H(z)}, \,\,\, \alpha_{\perp} = \frac{D_A(z)}{D^{\rm fid}_A(z)}
\end{eqnarray}
Hence the observed wavenumbers are given by $k^{\rm obs}_{\parallel} =  \alpha_{\parallel} k^{\rm true}_{\parallel}$, and 
$k^{\rm obs}_{\perp} = \alpha_{\perp} k^{\rm true}_{\perp}$. 
Including this scaling in Eq (\ref{eqn kaiser + damping}) one finds 
\citep{1996MNRAS.282..877B,1996ApJ...470L...1M,2010PhRvD..81d3512S}
\begin{eqnarray}
P_{\rm g}^s(k',\mu') &=& \frac{b^2}{\alpha_\perp^2 \alpha_\parallel} \left[ 1 +
  \mu'^2 \left( \frac{1+\beta}{\alpha_{\parallel}^2/ \alpha_{\perp}^2} - 1 \right)
  \right]^2 \nonumber \\ &\times& \left[ 1 + \mu'^2 \left(
  \frac{\alpha_{\perp}^2}{\alpha_{\parallel}^2} -1 \right) \right]^{-2} 
\\ &\times& P_{\rm \delta \delta} \left[ \frac{k'}{f_\perp} \sqrt{1 + \mu'^2
    \left( \frac{\alpha_{\perp}^2}{\alpha_{\parallel}^2} - 1 \right)} \right] \times D(\mu,k) \nonumber
\label{eqn:pkAP}
\end{eqnarray}
where $k'= \sqrt{(k^{\rm obs}_\perp)^2 + (k^{\rm obs}_\parallel)^2}$, $\mu' = k^{\rm obs}_\parallel/k'$, and $\beta = f/b$. 
This scaling introduces a new source of anisotropy in the clustering of galaxies, making it partially degenerate with redshift-space distortion
effects, accordingly it is important to account for this effect in this type of analysis
\citep{2012MNRAS.425..405B,2014MNRAS.443.1065B}.


Two components must be included to compare our theoretical predictions with observations: the window function and integral constraint effect, both of which result in   
a distortion to the measured 
power spectrum relative to the true power spectrum. Window function effects are induced by
the complex geometry of the survey (viz, a non-cubical geometry); and 
the integral constraint effect occurs as the condition $\delta_{k=0}=0$ is applied to the 
data: this imposed normalization for the $k=0$ mode is invalidated by 
super-survey modes.
Both effects induce a suppression of power at low--$k$ \citep{1991MNRAS.253..307P,2014MNRAS.443.1065B}.

A consistent comparison between our model and the observations therefore requires us to
include the window function effects in our modelling. 
Following \citet{2014MNRAS.443.1065B} the convolved multipoles $P_{l}^{\rm conv}(k)$ are 
calculated for the CMASS sample as
\begin{equation}
P^{\rm conv}_{\ell}(k) = 2\pi\int dk' k'^2\sum_L P^{\rm theory}_L(k')|W(k, k')|^2_{\ell L} - P_{l}^{ic}(k)\,,
\label{eq:conv}
\end{equation}
where
\begin{equation}
\begin{split}
|W (k,k')|^2_{\ell L} &= 2i^{\ell}(-i)^L(2\ell + 1)\sum^{N_{\rm ran}}_{ij, i\neq j}w_{\text{\tiny{FKP}}}(\vec{x}_i)w_{\text{\tiny{FKP}}}(\vec{x}_j)\\
&\;\;\;\;\;j_{\ell}(k|\Delta\vec{x}|) j_L(k'|\Delta\vec{x}|)\mathcal{L}_{\ell}({\hat{\vec{x}}_h\cdot \Delta\hat{\vec{x}}})\mathcal{L}_{L}({\hat{\vec{x}}_h\cdot \Delta\hat{\vec{x}}})\, , \nonumber
\end{split}
\label{eq:conv2}
\end{equation}
and the integral constraint term is given by
\begin{equation}
P^{\rm ic}_{\ell}(k) = 2\pi\frac{|W(k)|^2_{\ell}}{|W(0)|^2_{0}} \int dk'k'^2 \sum_LP_L^{\rm theory}(k')|W(k')|_L^2\frac{2}{2L+1} \,. \nonumber
\end{equation}
Here $j_L$ are spherical Bessel functions of order $L$, $N_{\rm ran}$ is the number of 
galaxies in the synthetic catalogue, and we sum over the monopole and quadrupole ($L = 0,2$).

Each survey region has a different window function and hence needs 
to be treated separately. 
To compute the CMASS likelihood we use 
the publicly available CMASS window 
functions\footnote{\url{https://sdss3.org/science/boss_publications.php}}.
The combined likelihood is now computed as
\begin{equation}
\begin{split}
-2\ln({\mathcal L}^{\rm WiggleZ}) - 2\ln({\mathcal L}^{\rm BOSS}) &=  \nonumber \\ 
\sum_{i = 1}^{12}(\vec{P}^{\rm WiggleZ}_{i} - \vec{P}^{\rm Conv}_{i})^{\rm T}&\hat{C}^{-1}_{{\rm Wig},i}(\vec{P}^{\rm WiggleZ}_{i} - \vec{P}^{\rm Conv}_{i}) \nonumber \\
\sum_{j = 1}^{2}(\vec{P}^{\rm BOSS}_{j} - \vec{P}^{\rm Conv}_{j})^{\rm T}&\hat{C}^{-1}_{{\rm BOSS},j}(\vec{P}^{\rm BOSS}_{j} - \vec{P}^{\rm Conv}_{j}) \, ,
\end{split}
\label{eq:like}
\end{equation}
The $i$ indices specify the two redshift bins ($z_{\rm eff} = 0.44, 0.73$) and six survey regions for WiggleZ ($12$ separate measurements).
The $j$ indices specify the two survey regions (NGC and SGC) for CMASS. Furthermore, 
$\vec{P}^{\rm WiggleZ}_{i} = [P_{0}^{\rm conv}(k),P_{2}^{\rm conv}(k),P_{4}^{\rm conv}(k)]_{\rm i}$ and
$\vec{P}^{\rm BOSS}_{j} =[P_{0}^{\rm conv}(k),P_{2}^{\rm conv}(k)]_{\rm j}$.

The hat in $\hat{C}^{-1}$ indicates that we are using a statistical estimator for the inverse covariance matrix. This estimator
is determined by the covariance matrix measured from mock catalogues: typically one would use $\hat{C}^{-1} = C^{-1}_{\rm mock}$, however,
the noise in the derived covariance matrix ($C^{-1}_{\rm mock}$) makes this estimator biased \citep{2007A&amp;A...464..399H}. We correct this bias 
using the estimator
\begin{equation}
\hat{C}^{-1} = \frac{N_s - n_b - 2}{N_s - 1}C_{\rm mock}^{-1},
\label{eq:covhartlap}
\end{equation} 
where $n_b$ is the number of power spectrum bins, and $N_{\rm s}$ the number of mock realisations used to construct the covariance matrix.

\subsection{BAOs}\label{BAOsect}

Acoustic oscillations in the photon-baryon plasma, prior to recombination,
imprint a series of fluctuations
in large-scale structure: in configuration-space one finds a
preference for galaxies to be distributed with a given comoving separation ($\sim 105$ \rUnit).

This excess in clustering (the BAO feature) functions as a cosmic yard-stick allowing
the cosmic expansion history to be mapped out. By
measuring the spherically averaged BAO position one determines
\begin{eqnarray}
D_{\rm V}(z) = \left[{cz(1 + z)^2D_{A}(z)^{2}}{H(z)}\right]^{1/3}.
\end{eqnarray}
Here $D_{A}(z)$ is the angular diameter distance. With higher signal to noise measurements one can
extract more information by isolating the transverse and line-of-sight BAO positions, determining
\begin{eqnarray}
\alpha_{\rm perp} &=& D_{\rm A}(z)r^{\rm fid}_{\rm s}/D^{\rm fid}_{A}(z)r_{\rm s}\\ 
\alpha_{\rm par} &=& H_{\rm fid}(z)r^{\rm fid}_{\rm s}/H(z)r_{\rm s} . 
\end{eqnarray}

By including the dependence on 
$r_{\rm s}$ (the sound horizon at the drag epoch), and expressing the measured quantity as a
ratio of the fiducial prediction, the dependence on CMB physics and the assumed cosmology has been 
made explicit.

To constrain the expansion history we use the following BAO measurements:
WiggleZ reconstructed from \cite{2014MNRAS.441.3524K},
reconstructed DR11--CMASS and DR11--LOWZ from \cite{2014MNRAS.441...24A}, and the
6dFGS measurement from \cite{2011MNRAS.416.3017B}. By `reconstructed' we are referring 
to the process of sharpening the acoustic peak by using information from the local 
density field \citep[cf.][]{Padmanabhan:2012oq}. 
The above measurements (excluding CMASS) can be incorporated into a likelihood given by
\be
-2\ln {\mathcal L} = (x - {\mathcal S})^{T}C^{-1}(x - {\mathcal S}) \, ,
\eeq
with the theory vector
\begin{eqnarray}
x &=& [ D_V(0.44)(r_{\rm fid}/r_{\rm s}), D_V(0.6)(r_{\rm fid}/r_{\rm s}),D_V(0.73)(r_{\rm fid}/r_{\rm s}) \nonumber \\
&& D_V(0.32)/r_{\rm d},r_{\rm s}/D_V(0.106)] \, ,
\end{eqnarray}
the data vector
\be
{\mathcal S} = [1716,2221,2516,8.25,0.336] \, ,
\ee
and the covariance matrix\footnote{We have scaled the WiggleZ elements for clarity; the true covariance 
matrix is obtained by scaling the WiggleZ elements by $ 10^{-4}$:
$C^{{\rm True} \; -1}_{\rm BAO \,(1,1)\;} =  2.17898\times 10^{-4}$.}
\begin{equation}
C^{-1}_{\rm BAO} = 
\begin{pmatrix} 
 2.17898 & −1.11633  & 0.46982 & 0 & 0 \\
−1.11633 & 1.70712  & −0.71847 & 0 & 0 \\
 0.46982 & −0.71847 & 1.65283 & 0 & 0  \\
0        & 0 & 0  & 36.025  & 0 \\
0        & 0 & 0  & 0  & 4444.4  \nonumber
\end{pmatrix}
\label{eq:likefinal2}
\end{equation} 
The CMASS measurements are in the form of probability distributions for
$P(\alpha_{\rm perp})$ and $P(\alpha_{\rm par})$ evaluated at $z_{\rm eff} = 0.57$. These measurements are therefore analysed 
separately, for details see \citet{2014MNRAS.441...24A}.
A number of these BAO measurement have been calculated using the approximate fitting formula for $r_{s}(z_{\rm d})$ from \cite{1998ApJ...496..605E}; hence throughout, where appropriate, the BAO measurements derived using this approximation are scaled to be consistent with the result from $\CAMB$ \cite[cf.,][]{2012MNRAS.427.2168M}.

To further improve the redshift range of our expansion history measurements we extend this `base' sample by including 
the Lyman-$\alpha$ BAO measurement from \cite{2014arXiv1404.1801D}, and 
the Quasar-Ly$\alpha$ cross-correlation measurement from \cite{2014JCAP...05..027F}.
The measurements are
$D_{\rm H}(z = 2.34)/r_{\rm s} = 9.18 \pm 0.28 \,, D_{\rm A}(z = 2.34)/r_{\rm s} = 11.28 \pm 0.65\,, 
D_{\rm H}(z = 2.36)/r_{\rm s} = 9.0 \pm 0.3\,, D_{\rm A}(z= 2.36)/r_{\rm s} = 10.8 \pm 0.4\, ,$
where $D_{\rm H } = c/H$.
Both common cosmic variance 
or a common source for the measurement error would induce correlations between the Lyman-$\alpha$ measurement. 
Fortunately, the origin of the dominant error components 
for these measurements are distinct, and hence the measurements are uncorrelated \citep{2014JCAP...05..027F}.
Additionally,
we treat any correlations between the BOSS and WiggleZ surveys as insignificant, 
given the small overlapping area ($ \sim 550$deg$^2$) and
the significance of shot noise in WiggleZ measurements.

\subsection{Growth Rate and Alcock-Paczynski Measurements}\label{Sec4: RSD}

The growth rate measurements presented in this section will be used to constrain $\gamma$.
Following the arguments presented in subsection \ref{subsec:modelPk} we only include growth rate constraints 
that have consistently incorporated the Alcock-Paczynski effect.
The exception to this point is for very low-redshift observations,
which are effectively insensitive to changes in the expansion history. 

In order to self-consistently express the degeneracy with the expansion history we 
chose to fit to joint $3$D posterior distributions from AP, BAO and RSD 
measurements: as opposed to marginalized $1$D constraints on $f\sigma_8(z)$.
The growth rate measurements we utilize are measured from BOSS-DR$11$ survey, 
the WiggleZ Dark Energy Survey,
and the $6$dF Galaxy survey \citep{2014MNRAS.443.1065B,2012MNRAS.425..405B,Beutler:2012fk}.
For the CMASS sample we use the data vector\footnote{This result is found fitting the power spectrum multipoles to $k_{\rm max} = 0.20$ \kUnit .}
\begin{eqnarray}
{\mathcal S}^{\rm BOSS}_{k_{\rm max}=0.20} &=& [D_V(0.57)/r_s(z_d), F_{\rm AP}(0.57), f(0.57)\sigma_8(0.57)] \nonumber \\ 
&=& [13.88, 0.683, 0.422] \, .
\label{eq:likefinal0}
\end{eqnarray} 
Where the AP effect translates into a geometric constraint on
$F_{\rm AP}(z) = (1+z)D_A(z)H(z)/c$. And the corresponding covariance matrix is given by 
\begin{equation}
10^3C_{k_{\rm max}=0.20}^{\rm BOSS} = \begin{pmatrix} 36.400 & -2.0636 & -1.8398\\
            0        & 1.0773 & 1.1755\\
            0        & 0 & 2.0438\end{pmatrix}
\label{eq:likefinal3}
\end{equation} 
The WiggleZ survey measurements are performed within three overlapping, hence correlated, 
redshift bins at $z_{\rm eff} = 0.44, 0.60, 0.73$. We first split the data vector into redshift bins, namely
${\mathcal S}^{\rm WiggleZ}_{k_{\rm max}=0.30} = ({\mathcal S}_{z_1},{\mathcal S}_{z_2},{\mathcal S}_{z_3})$.
In each of these redshift bins \citet{2012MNRAS.425..405B} measure the parameter combination
\begin{eqnarray}
{\mathcal S}_{\rm z_i} = \left[ A(z_i), F_{\rm AP}(z_i), f(z_i)\sigma_8(z_i) \right] \, ,
\label{eq:likefinal0new}
\end{eqnarray} 
where $A(z)$, the acoustic parameter, is given by
\begin{eqnarray}
A(z) \equiv \frac{100 D_{\rm V}(z) \sqrt{\Omega_{\rm m}h^{2}}}{cz} \, .
\end{eqnarray}
The measured values are now ${\mathcal S}_{\rm z_1} = (0.474,0.482,0.413)$, ${\mathcal S}_{\rm z_2} = (0.442,0.650,0.390)$, and 
${\mathcal S}_{\rm z_3} = (0.424,0.865,0.437)$. Table $2$ in \cite{2012MNRAS.425..405B} gives the full covariance matrix for ${\mathcal S}^{\rm WiggleZ}$. 

The final measurement we use is $f(0.106)\sigma_8(0.106) = 0.423 \pm 0.55$ 
from \cite{Beutler:2012fk}. As noted previously, the AP effect is not significant 
for this measurement given the low-redshift nature of the sample. 
All of the introduced measurements are now incorporated using the likelihood
\be
-2\ln {\mathcal L} = (x - {\mathcal S})^{T}C^{-1}(x - {\mathcal S}) \, ,
\eeq
here $x$, ${\mathcal S}$ and $C$ are the appropriate theory vector, data vector and covariance matrix. Note
that BAO information is included in both Sec. \ref{BAOsect} and Sec. \ref{Sec4: RSD} and we do not 
double-count this information.

\section{Secondary Datasets}\label{Sec3:sec_datasets}

A brief introduction and motivation is given for the additional datasets we use.

\subsection{Type-Ia SNe}

Sample variance effectively imposes a minimum volume limit for BAO detection. Accordingly, 
large volumes and hence higher redshift observations are preferable. Type-Ia SNe measurements 
do not have this restriction and hence can provide very accurate constraints on
the low-redshift expansion rate: an epoch where the presence of ``dark energy''
appears to dominate.

Therefore we include the distance modulus measurements for 
$473$ type Ia SNe presented in \citet{2011ApJS..192....1C}. The "SNLS" sample is
a combination of a number of previous surveys combining supernova legacy survey results with other 
low-$z$ and high-$z$ observations. These measurements are included in our 
analysis using the {\tt cosmomc} likelihood module provided by 
\citet{2011ApJS..192....1C}\footnote{ https://tspace.library.utoronto.ca/handle/1807/25390}.
This likelihood is evaluated by (firstly) calculating the model apparent magnitudes
(or more accurately, the rest-frame peak B-band magnitude):
\begin{eqnarray}
m_{\rm model} = 5 \log_{10} {\mathcal D_{\rm L}}(z_{\rm CMB}, z_{\rm Hel},\dots) - \alpha({\mathcal S} -1) + \beta {\mathcal C} + {\mathcal M}_{\rm B} \nonumber \, .
\end{eqnarray}
Here ${\mathcal D_{\rm L}}$ is luminosity distance with the dependence on the Hubble constant removed (it's dimensionless).
And $z_{\rm CMB}$ and $z_{\rm Hel}$ are the CMB frame and heliocentric frame redshifts of the SN. ${\mathcal M}_{\rm B}$
is a parameter which controls the zero-point and is a function of both the absolute magnitude
of the SN and $H_{0}$, this parameter is marginalised over.
The brightness of each 
SN is `standardised' using observations of the shape of the light curve, $s$, and the colour ${\mathcal C}$; in addition to the 
empirical relationship of these parameter with the luminosity of the object: these dependences are characterised by the 
parameters $\alpha$ and $\beta$.

Writing the model predictions as a vector $\vec{m}_{\rm model}$ the likelihood is given by
\be
-2\ln {\mathcal L} = (\vec{m}_{\rm obs} - \vec{m}_{\rm model})^{T}C^{-1}(\vec{m}_{\rm obs} - \vec{m}_{\rm model}) \, ,
\eeq
where $\vec{m}_{\rm obs}$ is a vector of the observed B-band magnitudes. The elements of the non-diagonal covariance matrix $C$ 
includes contributions from the following effects: the intrinsic-scatter 
of type Ia SN, the errors on the fitted light curve parameters, the redshift error, a host correction error, and  
the covariance between $s$, ${\mathcal C}$ and $m_{\rm obs}$. There are additional corrections for the local peculiar velocity 
field, for further details see \citet{2011ApJS..192....1C}.

In Section \ref{sec: par 2} we adopt a second SNe dataset, namely the JLA sample \citep{2014A&amp;A...568A..22B}.
This sample is composed of recalibrated SN Ia light-curves and distances for 
the SDSS-II and SNLS samples; this sample can be distinguished from the SNLS sample by the treatment of systematic affects, the end result is a $1.8\sigma$ shift from the SNLS 3-year results.

\subsection{CMB}

For the models we adopt GR is restored at the time of the last scattering surface; accordingly,
the components of the temperature fluctuations, unmodified by large-scale structure, 
provide a powerful tool to both constrain the physical components of the universe and 
the initial conditions which seed large-scale structure. 

The likelihood code for the power spectrum $C^{T T}_{l}$ from {\it Planck} is a hybrid: 
it is divided into high-$l$ and low-$l$.
For high-$l$ ($l>50$) we use the likelihood code {\tt CamSpec} described by \citet{2014A&amp;A...571A..15P}.
This algorithm uses temperature maps derived at 100, 143 and 217 GHz. Once both diffuse Galactic emission 
and Galactic dust emission are masked, $57.8$\% of the sky remains for the $100$ GHz map and $37.3$\% for 
the remaining maps.
At low multipoles ($2<l<49$) the likelihood is computed using the {\tt Commander} 
algorithm  \citep{2008ApJ...676...10E} using the frequency range 30–-353 GHz over
91\% of the sky.

Sub-Hubble modes near reionization are damped by Thomson scattering, thus obscuring our 
view of the primordial power spectrum: We observe a fluctuation amplitude $A_{s}e^{-2\tau}$. The 
degeneracy between the optical depth $\tau$ and the amplitude of the primordial power 
spectrum $A_{s}$ can be partially broken by including polarization data: the relative amplitude 
of the polarization and temperature power spectrum constrain $\tau$. For this purpose
we include the large-scale polarization measurements ($C^{E E}_{l}$) from 
{\it WMAP-9} \citep{2013ApJS..208...20B}. We use the likelihood code
from {\it Planck} which fits to the $l$-range ($2 < l < 32$). 

\subsubsection{CMB Lensing}

Photons travelling from the last scattering surface to our satellites encounter 
a number of over- and under-densities along the way. The intersected structure
deflects the photon paths and the large-scale clustering of matter causes
these deflection paths to be correlated over the sky \citep{1987A&amp;A...184....1B}. The 
combined effect of this CMB lensing is a re-mapping of the CMB 
temperature fluctuations \citep[cf.,][]{2006PhR...429....1L}:
\begin{eqnarray}
T(\hat{\bf n}) = T^{\rm unlensed}(\hat{n} + \nabla \Phi(\hat{\bf n})) \, . 
\end{eqnarray}
Where $\Phi(\hat{\bf n})$ is the CMB lensing potential given by
\begin{eqnarray}
\Phi(\hat{\bf n}) = -\int^{\chi*}_{0} d\chi \, G(\chi,\chi^*)\left[ \phi ( \chi \hat{\bf n};\eta_{0} - \chi) + \psi ( \chi \hat{\bf n};\eta_{0} - \chi)\right] \, .
\end{eqnarray}
Here $\chi$ is the conformal 
distance, $\eta$ is the conformal time ($\eta_{0}$ is the time today), 
and $G(\chi,\chi^*)$ is a weighting function.
The integration is taken from the last scattering surface ($\chi^*$) to today ($\chi = 0$);
hence this term represents the integrated effect of structure on photon paths, or more accurately,
since we are interested in testing GR, the integrated effect of spatial and curvature perturbations.

The lensing power spectrum $C^{\phi \,\phi}_{l}$ can be extracted from CMB maps; here
we use the results from \citet{Collaboration:2013qf} for the $l$-range $40 < l < 400$ (with the bin size $\Delta l = 64$):
this $l$-range is chosen as it encompasses the majority of the lensing signal ($\sim 90\%$) and is likely less influenced by 
systematic effects \citep[cf.,][]{Collaboration:2013qf}. Given the 
lensing kernel peaks at $z \sim 2$ and we are only using $l< 400$, the 
lensing power spectrum measurements used are only probing linear scales. Accordingly, 
we use linear theory to predict the lensing power spectrum and expect no 
systematic errors to be introduced from this modelling.

\subsubsection{Temperature-Galaxy Cross-Correlation}

At late times the accelerating cosmic expansion dictates the evolution 
of density perturbations, one consequence is
time-dependent metric potentials. This time-dependence is
apparent in the CMB as it generates a net energy loss for CMB photons 
as they propagate through these potential wells \citep{1967ApJ...147...73S}. This 
feature is known as the integrated Sachs-Wolfe (ISW) effect. The influence on 
the CMB power spectrum is given by
\begin{equation}
C_{l}  \sim ({\dot \phi} + {\dot \psi})\, .
\end{equation}
The ISW effect induces a correlation between the CMB (low--$l$) and 
large-scale structure probes: this is measured using the 
temperature-galaxy cross-correlation power spectrum $C^{g T}_{l}$ \citep[cf.][]{2008PhRvD..78d3519H}.  
For our analysis we use the measurement of $C^{g T}_{l}$ presented in \citet{2008PhRvD..78d3519H}, and 
the likelihood code described in \citet{2011PhRvD..84l3001D}. This likelihood code 
expands on that presented in \citet{2008PhRvD..78d3519H} by including the effects of 
modified gravitational field equations.

The density field for the cross-correlation is approximated by the following measurements: 
the $2$MASS Two Micron All Sky Survey, the Sloan-Digital Sky Survey Luminous Red Galaxy Sample,
the Sloan-Digital Sky Survey Quasars, and the NRAO VLA Sky Survey.
And the CMB temperature data is
taken from WMAP-5\footnote{Note, the NVSS radio survey is the best tracer of large-scale structure at a high-redshift: this survey 
provides the most significant detection of a cross-correlation. Furthermore,
the ISW effect is only dominant at low-$l$ and hence is limited by cosmic variance. For both reasons, the
measurement of $C^{g\,T}_{l}$ has not been significantly improved from \citet{2008PhRvD..78d3519H}, hence justifying our 
use of this data.}. 
The final $l$-range we adopt is $6<l<130$: this range is taken to ensure linear theory is valid,
specifically, this $l$-range is imposed to ensure a wavenumber cutoff of $k \le 0.05$ \,\kUnit.

\section{MCMC Analysis}\label{Sec3:MCMC analysis}

We sample the parameter space of cosmological parameters
using Markov Chain Monte Carlo techniques with the $\COSMOMC$ package. 
The MCMC algorithm implemented 
within this code is an adaptive Metropolis-Hastings method which  
utilizes a number of techniques to ensure fast convergence times. 
The definitions and adopted priors of each parameter are given in
Table \ref{tab:params}.
Our results are derived using $8$
separate chains which are run until convergence is achieved. 
The convergence of the Markov chains is determined
using the Gelman and Rubin convergence criteria, for which
chains require $R - 1 < 0.02$ to be satisfied for the least-converged orthogonalized parameter; 
$R$ being the ratio of the variance of the chains' mean and the mean of the chains' variances \citep{GR92}.
The posterior mean and $68\%$ confidence intervals are then computed using thinned Markov chains. 

There is currently no consensus on the $H_{0}$ value as measured from Cepheid data.
The most up-to-date measurements are presented by
\citet{2014MNRAS.440.1138E}, \citet{Riess:2011tg}, and \citet{2013arXiv1307.6031H}: 
they measure $H_{0} = 70.6 \pm 3.3, 73.8 \pm 2.4, 72.0 \pm 3$ km/s/Mpc, respectively.
Note both \citet{2014MNRAS.440.1138E} and \citet{2013arXiv1307.6031H} have used the revised
geometric maser distance to NGC 4258 \citep[as presented in][]{2013arXiv1307.6031H}, however their
measurements still do not agree: the disagreement can be traced to
different outlier rejection criteria being applied.
For this analysis we adopt two approaches, because of this tension. When the expansion 
history is described by $\Lambda$CDM we do not include any $H_{0}$ prior as the model-dependent 
constraints from the CMB are sufficient. When we do include deviations from 
$\Lambda$CDM in the expansion history we add an $H_{0}$ prior using the measurement by \citet{2014MNRAS.440.1138E}.

\begin{table*}
\begin{center}
\begingroup
\newdimen\tblskip \tblskip=5pt
\caption{Cosmological parameters used in our analysis. For each we give the symbol, uniform prior range, value taken in the \lcdm\ cosmology, and summary definition. The parameters with a specified prior range are treated as free parameters in the MCMC analysis, 
while the remaining parameters are fixed at their fiducial values. The first block contains the standard parameters present in the 
\lcdm\ model, while the second and third contain the parameters introduced to allow modifications from General Relativity. Note a prior is included on the derived parameter $H_{0}$.}
\label{tab:params}
\vskip -5mm
\footnotesize 
\setbox\tablebox=\vbox{ %
\newdimen\digitwidth 
\setbox0=\hbox{\rm 0}
\digitwidth=\wd0
\catcode`*=\active
\def*{\kern\digitwidth}
\newdimen\signwidth
\setbox0=\hbox{+}
\signwidth=\wd0
\catcode`!=\active
\def!{\kern\signwidth}
\halign{\hbox to 2.7cm{#\leaderfil}\tabskip=0.4cm& \hfil#\hfil\tabskip=0.6cm&
 \hfil#\hfil\tabskip=0.6cm&  #\hfil\tabskip=0pt\cr
\noalign{\doubleline}
\omit\hfil Parameter\hfil&\omit\hfil Prior range\hfil&\omit\hfil Baseline\hfil&\omit\hfil Definition\hfil\cr
\noalign{\vskip 3pt\hrule\vskip 3pt}

$\omb \equiv \Omb h^2$& $[0.005, 0.1]$ & \dots& Baryon density today\cr
$\omc \equiv \Omc h^2$& $[0.001, 0.99]$& \dots& Cold dark matter density today\cr
$100\theta_{\mathrm{MC}}$ & $[0.5, 10.0]$ & \dots& $100\,{\times}$ approximation to $\rstar/D_{\rm A}$ \cr
$\tau                $&   $[0.01, 0.8]$ & \dots& Thomson scattering optical depth due to reionization \cr
$\ns           $& $[0.9, 1.1]$ & \dots& Scalar spectrum index ($k_0 = 0.05\Mpc^{-1}$)\cr
$\ln(10^{10}\As) $& $[2.7, 4.0]$ & \dots& Log power of the primordial curvature perturbations ($k_0 = 0.05\,\Mpc^{-1}$)\cr
$\Omk            $&   & 0& Curvature parameter today \cr
$\mnu        $& & $0.06$ & The sum of neutrino masses in eV \cr
$\neff       $& & $3.046 $ & Effective number of relativistic degrees of freedom \cr
\noalign{\vskip 3pt\hrule\vskip 3pt}
{\it Parameters model} {\color{red}\rom{1}}            &   & & \cr
$z_{t}                $ &    & $1.0$ & Transition redshift for GR modifications \cr
$k_{t}                $ &    & $0.01$ & Transition wavenumber for GR modifications ($\Mpc^{-1})$ \cr 
$\gl(k,z)                $&   $[-10,10]$ & $1$ & Modification to relativistic Poisson equation (Eq \ref{eqn:GR modifications})  \cr
$\gm(k,z)                $&   $[-10,10]$ & $1$ & Modification to non-relativistic Poisson equation (Eq \ref{eqn:GR modifications}) \cr
\noalign{\vskip 3pt\hrule\vskip 3pt}
{\it Parameters model} {\color{red}\rom{2}}          &   & & \cr
$w_0                $& $[-3.0, 1.0]$ & $-1$& Dark energy equation of state, $w(a) = w_0 + (1-a) w_a$\cr
$w_a                 $& $[-3, 3]$ & 0& Redshift-dependent modification to the equation of state (see above) \cr
$\gamma$              & $[0,2]$ & $0.55$& Power-law index of the growth-rate parameter $f(z)=\Omega_{\rm m}^{\gamma}$ \cr
\noalign{\vskip 3pt\hrule\vskip 3pt}
$\Oml      $&     & \dots& Dark energy density divided by the critical density today\cr
$\Omm     $&  & \dots& Matter density today divided by the critical density\cr
$\sigma_8            $&   & \dots& RMS matter fluctuations today in linear theory\cr
$H_0                 $&[20,100] & \dots&  Expansion rate today in $\rm{km}\, \rm{s}^{-1}\Mpc^{-1}$\cr
\noalign{\vskip 3pt\hrule\vskip 3pt}}}
\endPlancktablewide 
\endgroup
\end{center}
\end{table*}

\subsection{Parameter Fits: Model {\color{red}\rom{1}}}

Using different combinations of the measurements outlined in the previous sections, we 
performed fits to the base $\Lambda$CDM parameters 
($\omb,\omc,\theta_{\mathrm{MC}},\tau,\ns,\As$) and the modified gravity parameters 
$\gm(k,z)$ and $\gl(k,z)$. Recall each modified gravity parameter is binned in both redshift and 
scale.

In addition to the physical parameters, a number of nuisance parameters
are introduced to account for unknown astrophysical effects. For the WiggleZ multipole 
calculation for each redshift bin we include the galaxy bias and velocity dispersion as 
nuisance parameters, that is,
$b_{\rm lin}(z=0.44),\sigma_v(z=0.44), b_{\rm lin}(z=0.73)$, and $\sigma_v(z=0.73)$.
The uniform priors imposed on these parameters are $b_{\rm lin} \in [0.5,3]$ and 
$\sigma_v \in [0,10]$\rUnit. For the DR$11$-BOSS CMASS multipole measurement we also include 
galaxy bias and velocity dispersion as free parameters, $b_{\rm lin}(z= 0.53),\sigma_v(z=0.53)$.
Additionally for BOSS, we include a free parameter
to account for the shot noise $N$, this is given the prior 
$ N \in [0,2000]$\vUnit. For the 
WiggleZ measurement the shot noise contribution has already been subtracted.
For the velocity power spectrum measurement we include a velocity dispersion parameter 
$\sigma_{\rm PV}(z=0) \in [0,500]{\rm km/s}$.

\begin{figure*}
\centering
\includegraphics[width=16cm]{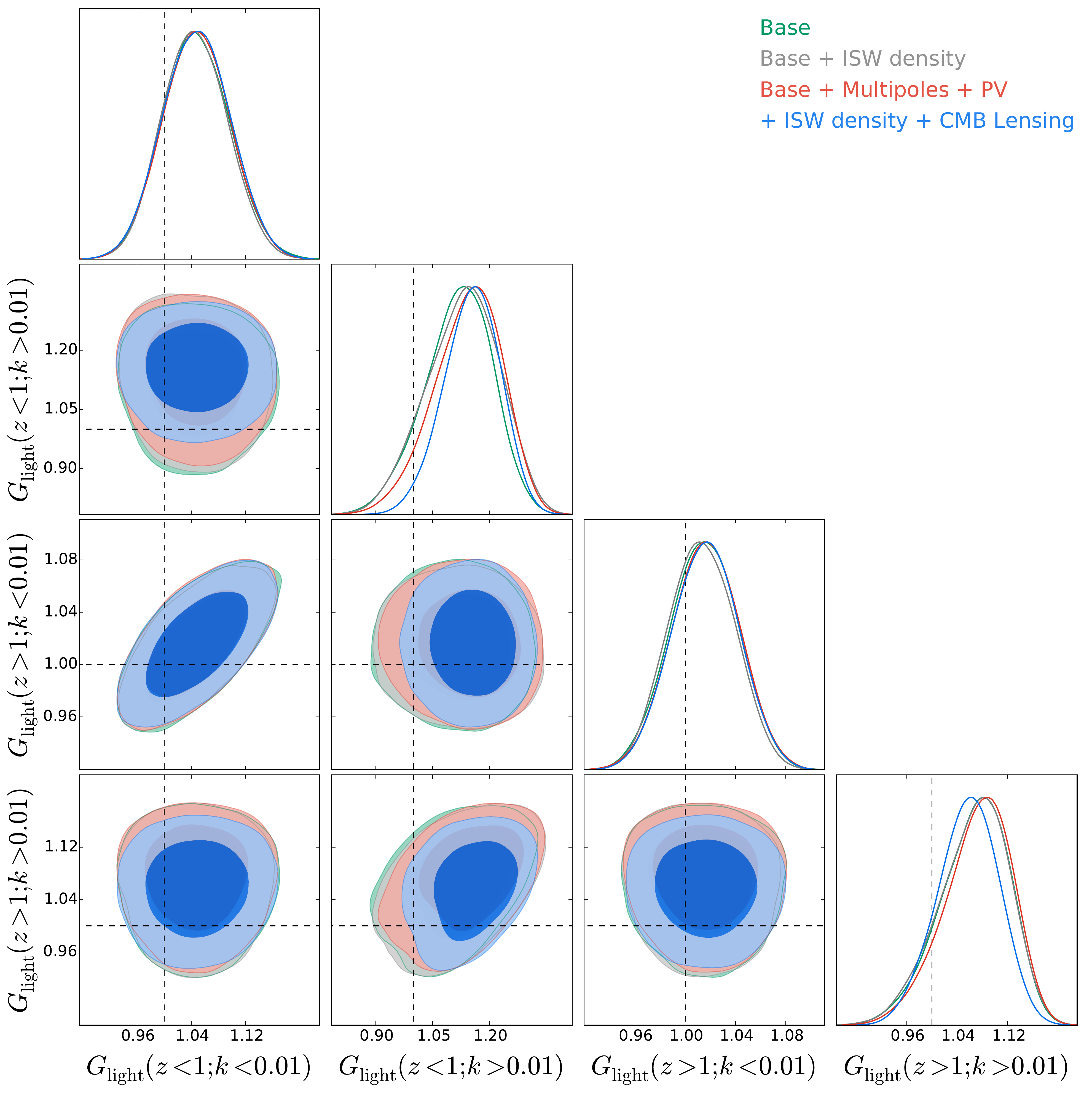}
\caption{$68$\% and $95$\% confidence regions for the four $\gl(k,z)$ bin parameters. 
Here $z>1$ is referring to the redshift range $2>z>1$. 
Note all of the parameters specified in Table 2 are being varied in this analysis, however for clarity we 
only plot the constraints on $\gl(k,z)$ in this plot. Recall we have defined 
Base as ${\rm High-}l + {\rm low-}l + {\rm WP} + {\rm BAO} + {\rm SNe}$. }  
\label{MCMC_G_light}
\end{figure*}

\begin{figure*}
\centering
\includegraphics[width=16cm]{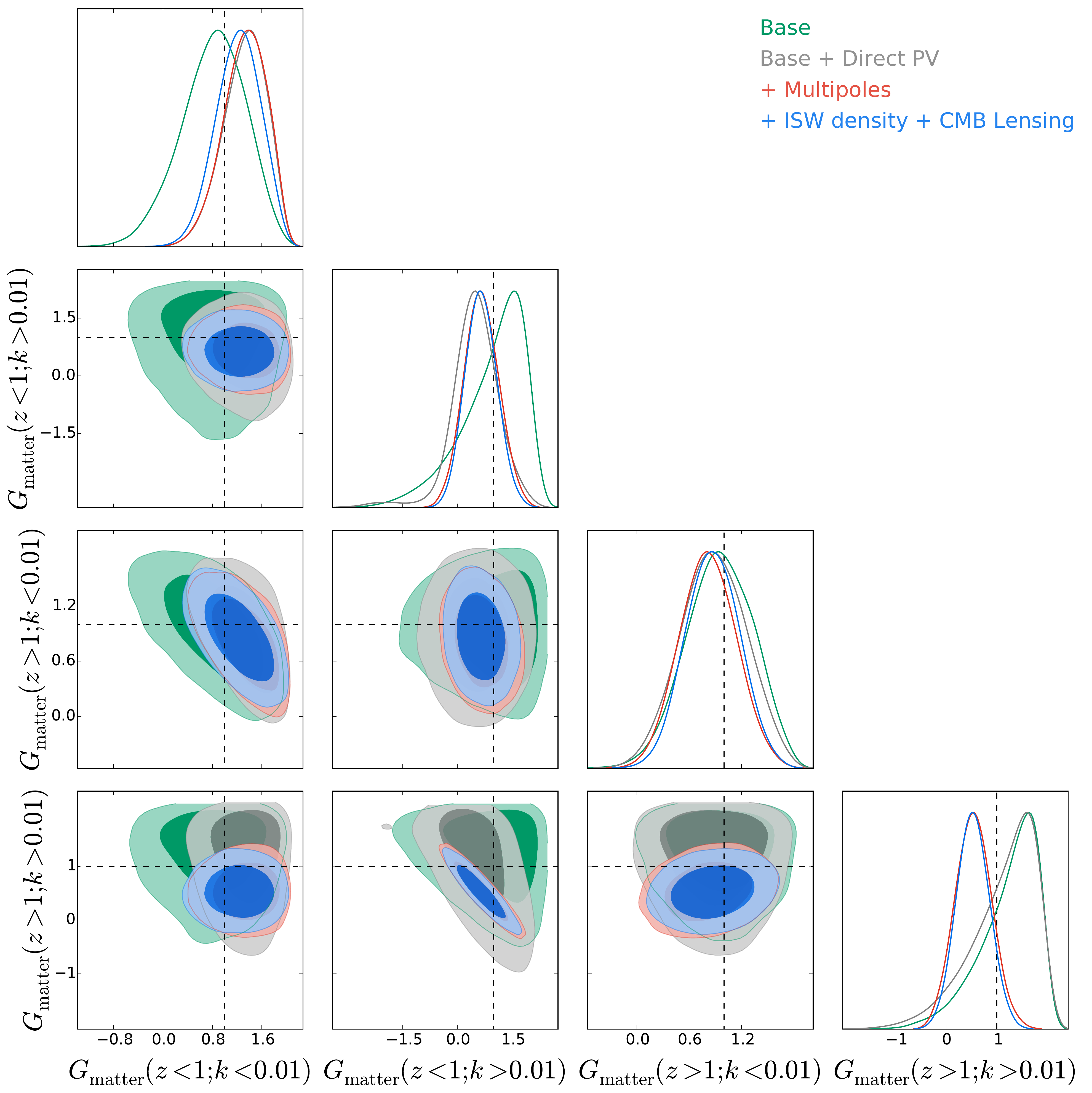}
\caption{$68$\% and $95$\% confidence regions for the four $\gm(k,z)$ bin parameters.
Here $z>1$ here is referring to the redshift range $2>z>1$.  
Note all of the parameters specified in Table 2 are being varied in this analysis yet for clarity we 
only plot the constraints on $\gm(k,z)$. Recall we have defined 
Base as ${\rm High-}l + {\rm low-}l + {\rm WP} + {\rm BAO} + {\rm SNe}$.}
\label{MCMC_G_matter}
\end{figure*}

In order to understand the sensitivity of each cosmological probe
to the physical parameters, and test for residual systematics, we 
analyse different combinations of cosmological probes. 
The different combinations are defined and labeled in Table \ref{tab:datasetsMod1}
(henceforth we will use these definitions).
The final results of this section are displayed in Fig. \ref{MCMC_G_light} and Fig. \ref{MCMC_G_matter}, and
further information is provided in Table \ref{MCMC_main_table}. The first figure shows 
the constraints on $\gm(k,z)$ and the second on $\gl(k,z)$. The black-dashed lines 
in both figures show the predictions from General Relativity.
We do not plot the 
2D contours between $\gm(k,z)$ and $\gl(k,z)$ as their correlations are 
small, i.e., $\langle |\rho_{c}| \rangle \sim 0.15$. Here $\rho_{c}$ is the 
cross-correlation coefficient, and $\langle \, \rangle$ indicates the average over
all the possible values between $\gm(k,z)$ and $\gl(k,z)$. 
Similarly, we do not plot the inferred constraints on the base $\Lambda$CDM parameters
as, with two exceptions, the base $\Lambda$CDM parameters are not highly 
correlated with the post-GR parameters, the exception being $\sigma_{8}$ and $\Omega_{\rm m}$ 
with $G_{\rm matter}$. When averaging over the four $G_{\rm matter}$
parameters we find $\langle |\rho_{c}| \rangle \sim 0.77,0.39$, respectively.
The remainder of this section will 
involve a discussion of the content of these plots, in addition to some comments 
on potential systematics effects and the derived astrophysical parameters.

\begin{table}
  \caption{The dataset combinations we use for fits to Model {\color{red}\rom{1}}, in addition to 
  the labels we adopt to refer to them. The corresponding datasets should be clear 
  from the information given in Table \ref{tab:datasets}. We define 
 Base as  the combination $ {\rm High-}l + {\rm low-}l + {\rm WP}
   + {\rm BAO} + {\rm SNe}$. Below {\tt CMASS} refers to the monopole and quadrupole 
   multipole measurements from the BOSS-CMASS sample. And {WiggleZ} refers to the 
   monopole, quadrupole and hexadecapole measurements from WiggleZ (as presented above).}
\begin{tabular}{ll}
\noalign{\doubleline}
 \multicolumn{1}{c}{Label} & \multicolumn{1}{c}{Description} \\
\noalign{\vskip 3pt\hrule\vskip 3pt}
{\tt SET} 1   &  Base  \\ [0.6mm] 
{\tt SET} 2   &  Base + Direct PV \\ [0.6mm] 
{\tt SET} 3   &  Base + CMASS ($k_{\rm max} = 0.10$\kUnit)\\
&  WiggleZ ($k=0.15$\kUnit) + Direct PV  \\[0.6mm]   
{\tt SET} 4   &  Base + CMASS ($k_{\rm max} = 0.10$\kUnit) + \\
& WiggleZ ($k=0.15$\kUnit) +  Direct PV \\
& + ISW-Density + CMB Lensing \\ [0.6mm]  
{\tt SET} 5   &  Base  + ISW-Density \\ [0.6mm]   
{\tt SET} 6   &  Base  + CMASS ($k_{\rm max} = 0.10 $\kUnit) \\ [0.6mm] 
{\tt SET} 7   &  Base + CMASS ($k_{\rm max} = 0.15$\kUnit)    \\[0.6mm]
{\tt SET} 8   &  Base + WiggleZ ($k_{\rm max} = 0.15$\kUnit)   \\[0.6mm]   
{\tt SET} 9   &  Base +  WiggleZ ($k_{\rm max} = 0.19$\kUnit)  \\[0.6mm] 
\hline
\end{tabular}
\label{tab:datasetsMod1}
\end{table}

\begin{table*} 
\begin{center}
\caption{Cosmological parameter constraints for Model {\color{red}\rom{1}}. The constraints are derived from four different 
groups of cosmological probes, we labels these groups Set 1 to 4 and define each in Table \ref{tab:datasetsMod1}. 
For each parameter in each group we provide the $68\%$ confidence levels. 
To keep the table a reasonable size we only consider the parameters most relevant to our analysis.
}
\begingroup
\openup 2pt
\newdimen\tblskip \tblskip=5pt
\nointerlineskip
\vskip -3.0mm
\footnotesize
\setbox\tablebox=\vbox{
    \newdimen\digitwidth
    \setbox0=\hbox{\rm 0}
    \digitwidth=\wd0
    \catcode`"=\active
    \def"{\kern\digitwidth}
    \newdimen\signwidth
    \setbox0=\hbox{+}
    \signwidth=\wd0
    \catcode`!=\active
    \def!{\kern\signwidth}
\halign{
\hbox to 0.9in{$#$\leaderfil}\tabskip=3.5em&$#$\hfil&$#$\hfil&$#$\hfil&\hfil$#$\hfil\tabskip=0pt\cr
\noalign{\doubleline}
\multispan1\hfil \hfil&\multispan1\hfil SET 1 \hfil&\multispan1\hfil SET 2 \hfil&\multispan1\hfil SET 3 \hfil&\multispan1\hfil SET 4 \hfil\cr
\noalign{\vskip -3pt}
\omit&\multispan1\hrulefill&\multispan1\hrulefill&\multispan1\hrulefill&\multispan1\hrulefill\cr
\omit\hfil Parameter\hfil&\omit\hfil 68\% limits\hfil&\omit\hfil 68\% limits\hfil&\omit\hfil 68\% limits\hfil&\omit\hfil 68\% limits\hfil\cr
\noalign{\vskip 3pt\hrule\vskip 5pt}
G_{\rm matter}(z<1;k>0.01)& 0.96^{+1.1}_{-0.44}     & 0.48^{+0.59}_{-0.52}          & 0.66\pm 0.47              & 0.65\pm 0.43       \cr
G_{\rm matter}(z<1;k<0.01)& 0.81^{+0.59}_{-0.46}     & 1.32^{+0.42}_{-0.29}         & 1.32^{+0.41}_{-0.30}      & 1.22^{+0.39}_{-0.34}      \cr
G_{\rm matter}(z>1;k>0.01)& 1.23^{+0.71}_{-0.28}     & 1.12^{+0.81}_{-0.33}         & 0.54\pm 0.35              & 0.53\pm 0.32       \cr
G_{\rm matter}(z>1;k<0.01)& 0.95^{+0.42}_{-0.36}    & 0.88\pm 0.37                  & 0.82\pm 0.32              & 0.87\pm 0.30       \cr
G_{\rm light}(z>1;k>0.01)& 1.067^{+0.063}_{-0.046}     &1.066^{+0.064}_{-0.045}      &1.072^{+0.063}_{-0.043}   &1.057^{+0.053}_{-0.045}       \cr    
G_{\rm light}(z<1;k<0.01)& 1.048\pm 0.048     & 1.044\pm 0.050                       &1.048\pm 0.048            &1.048\pm 0.048       \cr
G_{\rm light}(z<1;k>0.01)& 1.12^{+0.10}_{-0.078}     & 1.113^{+0.098}_{-0.084}       &1.14^{+0.10}_{-0.077}     & 1.153^{+0.080}_{-0.068}      \cr
G_{\rm light}(z>1;k<0.01)& 1.015\pm 0.026     & 1.016\pm 0.027                       &1.016\pm 0.026            & 1.016\pm 0.026       \cr
\Omega_b h^2& 0.02228\pm 0.00025    &0.02227\pm 0.00025                              &0.02226\pm 0.00025        & 0.02230\pm 0.00025       \cr
\Omega_c h^2& 0.1172\pm 0.0013       &0.1172\pm 0.0013                               &0.1168\pm 0.0013          & 0.1163\pm 0.0013       \cr
\tau& 0.087\pm 0.013       &0.089^{+0.012}_{-0.014}                                  &0.089\pm 0.013            & 0.086\pm 0.012       \cr
{\rm{ln}}(10^{10} A_s)& 3.076\pm 0.026         & 3.081\pm 0.025                      &3.080\pm 0.025            & 3.073\pm 0.025       \cr
\Omega_\Lambda& 0.7011\pm 0.0077    & 0.7009\pm 0.0078                               &0.7031\pm 0.0074          & 0.7060\pm 0.0074       \cr
\Omega_m & 0.2989\pm 0.0077      & 0.2991\pm 0.0078                                  &0.2969\pm 0.0074          & 0.2940\pm 0.0074       \cr
\sigma_8&   0.851^{+0.14}_{-0.091}       &0.783^{+0.095}_{-0.063}                    &0.717^{+0.018}_{-0.022}   & 0.711^{+0.017}_{-0.020}       \cr
H_0&    68.49\pm 0.61        & 68.46\pm 0.63                                         &68.61\pm 0.59             &68.83\pm 0.61       \cr
\noalign{\vskip 5pt\hrule\vskip 3pt}
}
}
\endPlancktable
\endgroup\label{MCMC_main_table}
\end{center}
\end{table*}

As shown in Fig. \ref{MCMC_G_light}, we observe very little variation in $G_{\rm light}(k,z)$ as we 
add extra datasets to the base sample (the green contour):
this is because the ISW effect on the T-T power spectrum is dominating the fit; additionally, 
galaxy velocities have no sensitivity to $G_{\rm light}(k,z)$, so we expect the benefit of including them to be minimal.
The grey contours in Fig. \ref{MCMC_G_light} are derived by adding the T-g measurements to the base sample, and the red contours are derived 
by adding the multipole and velocity measurements to the base sample.
And the blue contours show the main results which are derived using Set 4.
From these measurements for $G_{\rm light}(k,z)$ we infer (in terms of 68\% CLs)
\begin{eqnarray}
G_{\rm light}(z>1;k>0.01)&=& 1.057^{+0.053}_{-0.045}  	\,, \nonumber  \\
G_{\rm light}(z<1;k<0.01)&=& 1.048\pm 0.048  	\,,  \nonumber\\
G_{\rm light}(z<1;k>0.01)&=& 1.153^{+0.080}_{-0.068}	\,, \nonumber\\
G_{\rm light}(z>1;k<0.01)&=& 1.016\pm 0.026	\,,  \nonumber 
\end{eqnarray}
These measurements are compatible at the $95\%$ CL with GR.

For $G_{\rm matter}(k,z)$ we observe a significant amount of variation as 
new measurements are added to the base sample. In Fig. \ref{MCMC_G_matter} the green, grey, red and blue contours 
correspond respectively to measurements using the dataset combinations Set 1, Set 2, Set 3, Set 4 
(sets 6 to 9 are used for systematics checks to be discussed in the next section). 
As derived from Set 4 (i.e. using all the datasets) the $1$D marginalised results for $\gm$ (in terms of $68\%$ CLs) are
\begin{eqnarray}
G_{\rm matter}(z<1;k>0.01)&=& 0.65\pm 0.43 	\, ,     \nonumber  \\
G_{\rm matter}(z<1;k<0.01)&=& 1.22^{+0.39}_{-0.34}	\, ,    \nonumber \\
G_{\rm matter}(z>1;k>0.01)&=& 0.53\pm 0.32 	\, , 	\nonumber \\
G_{\rm matter}(z>1;k<0.01)&=& 0.87\pm 0.30	\, . 	\nonumber 
\end{eqnarray}

Similarly to above, these results are consistent with GR at the $95$\% CL, while at the $68\%$ CL level we observe a
tension with GR in the high-redshift and large-wavenumber bin. Furthermore, the constraints from Set 4 on
the $2$D CLs of the low-$z$ high-$k$ and high-$z$ high-$k$ bins of $G_{\rm matter}$
show a tension with the standard model at greater than $2\sigma$.
For the $1$D marginalised results this tension is significantly reduced as the high-$z$ and low-$z$ $G_{\rm matter}$ bins 
are highly correlated, as can be seen in Fig. \ref{MCMC_G_matter}.
This degeneracy occurs as some probes, such as the CMB, are sensitive to integrated 
quantities over redshift, such that higher growth at high-$z$ can be compensated for by lower 
growth at low-$z$.

Introducing direct PV measurements the constraints shift from the green to the grey contours. 
The most prominent shift occurs in the low-$z$ and 
low-$k$ $G_{\rm matter}$ bin, as expected: we find a shift from $G_{\rm matter}(z<1;k<0.01) = 0.81^{+0.59}_{-0.46}$  
to $G_{\rm matter}(z<1;k<0.01) = 1.32^{+0.42}_{-0.29}$.
We find further improvements in the constraints for the high-wavenumber and low-redshift bin. 
Future PV surveys should be able to considerably improve on this situation \citep[cf.][]{2014MNRAS.445.4267K}.
Using the best-fit parameters from Set 4, we measure $\chi^{2}_{\, 6{\rm dFGSv}} = 778$ with
$979$ data points:
the full $6$dfGSv velocity field is smoothed onto a grid with $979$ non-empty elements \citep[cf.][]{2014MNRAS.444.3926J}.

Including RSD measurements results in the shift from the grey to red contours, for which
we find a significant improvement in the constraint on the high-$z$ and high-$k$ $G_{\rm matter}$ bin.
Moreover, we find that the RSD measurements have more influence on the high-$z$ bin
than the the low-$z$ bin: this is an 
further consequence of measuring integrated quantities. 
As a systematic check we isolate the measurements from WiggleZ 
and BOSS and perform separate fits, we find that
the two separate constraints on $\gm$ are consistent. We can also 
assess how well our model fits the observations. By adding the multipole 
likelihoods we find $\Delta \chi^{2} = 322$, for a total of $324$ measurement points.
Individually, for the fit to the WiggleZ multipoles, with $126$ data points per redshift
bin we measure $\chi^{2}_{\rm WiggleZ} = 129.88$ for the low-$z$ region, and 
$\chi^{2}_{\rm WiggleZ} = 121.6$ for the high-$z$ region. Finally, for BOSS, given we are fitting to $k_{\rm max}=0.10$\kUnit \,,
there are $72$ measurement points and we find $\chi^{2}_{\rm CMASS} = 72.6$. 
   
\subsubsection{Astrophysical Parameters and Systematic Checks}

When calculating the power spectrum multipole predictions, 
we assumed a linear bias factor and linear perturbation theory. The validity of both
assumptions may be questioned.
We examine, albeit crudely, the importance 
of these assumptions by determining the sensitivity of the parameter fits 
to the small-scale cut-off $k_{\rm max}$. 
For our model fits using the CMASS and WiggleZ multipole likelihood calculations
we ran new Markov chains using different cut-off values
$k^{\rm CMASS}_{\rm max}= 0.10,0.15$\kUnit
and $k^{\rm WiggleZ}_{\rm max}= 0.15,0.19$\kUnit. 
The results showed no statistically significant shift when the  
fitting range was changed. 

The astrophysical parameters for the multipole and direct PV fits only vary slightly when using different 
dataset combinations, hence we choose to only present results from 
Set 4 (given in terms of $68$\% CLs).
For the fit to the WiggleZ multipole we find  $\sigma_{\rm v}(z = 0.73) = 2.30^{+1.2}_{-1.8}\,$\rUnit,
$\sigma_{\rm v}(z = 0.44) = 4.468^{+1.8}_{-1.0}\,$\rUnit, $b_{1}(z = 0.44) = 1.089\pm 0.042$, and 
$b_{1}(z = 0.73) = 1.207\pm 0.059$. For the fit to CMASS we find 
$\sigma_{\rm v}(z=0.57) = 2.44^{+0.68}_{-1.2}\,$\rUnit, $b_{1}(z=0.57) = 2.055\pm 0.084$, and 
$N({\rm Shot \, Noise}) = 705 \pm 200$\,\vUnit. Finally, from the fit to the velocity power spectrum 
we determine the $95\%$ upper limit $\sigma_{\rm PV}(z=0) < 334.6$km/s. With 
different $k_{\rm max}$ values adopted, one should not necessarily compare our results for the shot noise and 
velocity dispersion with previous analysis; however,
we find our bias measurements to be consistent with previous analysis. 

\subsection{Previous Measurements: Summary and Comparisons}

Below we briefly summarise recent work in this field, with a focus 
on results that adopt a similar parameterisation.

\begin{itemize}

\item{ \citet{2010PhRvD..82j3523D} presented constraints on 
$\{\mathcal G, V\}$, our $\{G_{\rm light}, G_{\rm matter} \}$, 
in bins of time and wavenumber.
To constrain these parameters they used the following probes: 
WMAP7, supernova Union2, CFHTLS weak lensing data, temperature-galaxy cross correlation, 
and the galaxy power spectrum. They identify the CFHTLS 
survey as responsible for a $2\sigma$ tension with GR in the 
high-$k$ and low-$z$ bin for ${\mathcal V}$. This feature 
is not observed when using the COSMOS data or in subsequent 
analysis of the final CFHTLenS catalogue. Note RSD information was not included, therefore the 
final constraints on $V$ are of order $\sim1$.}

\item{ \citet{2013MNRAS.429.2249S} measure the parameters $\{ \Sigma,\mu\}$ (i.e., $\{G_{\rm light}, G_{\rm matter} \}$) using 
tomographic weak lensing measurements from CFHTLenS and RSD measurements of $f\sigma_{8}$ from 6dFGS and WiggleZ,
in addition to WMAP7 (including low-$l$) and geometric information (see also \citet{2015arXiv150103119D} and \cite{2012PhRvD..85l3546Z}).
Their measurements are consistent with GR: they find
$\mu = 1.05 \pm 0.25$ and $\Sigma = 1.00 \pm 0.14$. 
For this fit they assumed $\Sigma,\mu$ are scale-independent and adopt a specific functional form for their temporal
evolution: this effectively confines deviations to very low-redshifts. 
Measurements of the t-g cross correlations, CMB-lensing, and the growth rate measurement from CMASS 
were not included in these fits.}

\item{\citet{2015arXiv150201590P} have recently provided the 
state-of-the-art measurements of post-GR parameters,
placing constraints on a extensive range of specific and phenomenological models.
For the phenomenological model they adopt the parameters $\{\mu,\eta\}$, as implemented in {\tt MGCAMB}.  
Motivated by $f(R)$ models, a specific functional form for the redshift and scale dependence of 
these parameters is assumed. As appropriate to their aim, they
ensure their angular cuts to the tomographic shear-shear measurements from CFHTLenS isolate the linear signal (see their Fig 2).
This approach is not adopted throughout, however. Their adopted $f\sigma_{8}(z=0.57)$ measurement (by \citet{2014MNRAS.439.3504S}) was
derived by fitting to the monopole and quadruple of the correlation function on scales larger than $25$\rUnit. As highlighted 
by the authors (see their Fig. 7) non-linear terms are significant on these length scales, the result is a
dependence on non-linear physics.}

\end{itemize}

In relation to the most up-to-date measurements, our results can be distinguished in two main
ways: Firstly, the inclusion of the velocity power spectrum measurements, which improve low-$k$ constraints; secondly,
the methodology we use to analyse RSD measurements, and the range of RSD measurements analysed. 
We argue that the methodology of directly analysing the power spectrum multipoles 
allows constraints to be derived that are more widely applicable to non-standard cosmological models. 
This is because it allows one to restrict the analysis to scales within the linear regime, where 
the phenomenological model we use describe physical models (see Sect. \ref{Introduction}). 
Moreover, the multipoles contain scale-dependent information, which is necessary if 
scale-dependent terms are introduced.

\subsection{Parameter Fits: Model {\color{red}\rom{2}}}
\label{sec: par 2}

We now explore fits to a new parameter space
that is more rigid regarding the allowed deviation to the growth history. 
Two scenarios will be considered when fitting for these parameters, firstly, 
an expansion history fixed to $\Lambda$CDM; and secondly,  
an expansion history than can deviate from $\Lambda$CDM via. a time-dependent equation of state.
We define the two parameter spaces as 
${\bf p_1} =\{\gamma,\omb,\omc,\theta_{\mathrm{MC}},\tau,\ns,\As \}$, and
${\bf p_2} =\{\gamma,w,w_{a},\omb,\omc,\theta_{\mathrm{MC}},\tau,\ns,\As \}$.
We choose not to include the influence that deviations in the expansion history have 
on the expected growth (that is, the relation $\gamma = f(\gamma_{0},w_{0},w_{a})$) as
the corrections are currently small. 

Note that by changing the growth rate we modify $\sigma_{8}$, 
this effect is included by altering the growth history well into 
the matter dominated regime. The modified growth factor is calculated as
\begin{eqnarray}
D(a_{\rm eff}) = \exp{\left( -\int^{1}_{a_{\rm eff}}da \, \Omega_{\rm m}(a)^{0.55}/a \right)} \, ,
\end{eqnarray}
now we scale the fiducial prediction $\sigma_{8}^{\rm Fid}(z_{\rm high})$ to find the 
modified amplitude $\sigma_{8}^{\gamma}(z_{\rm eff})$: 
\begin{eqnarray}
\sigma_{8}^{\gamma}(z_{\rm eff}) = \frac{D(a_{\rm eff})}{D(a_{\rm high})} \sigma_{8}^{\rm Fid}(z_{\rm high}) \, .
\end{eqnarray}

\begin{figure*}
\centering
\includegraphics[width=16.0cm]{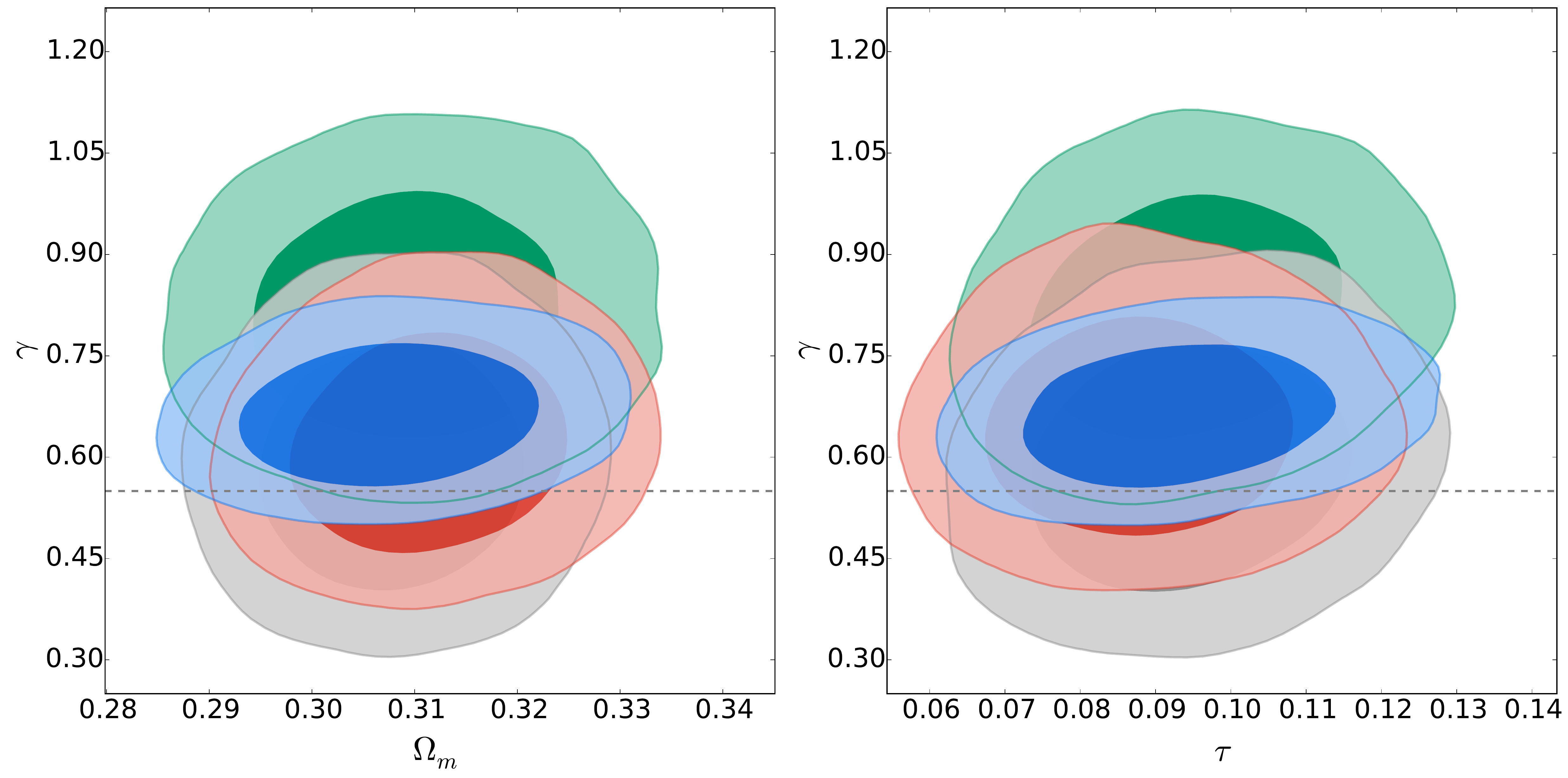}
\caption{$2$D marginalized posterior distributions for $\{\Omega_{\rm m},\gamma\}$ and $\{\tau,\gamma\}$, assuming a $\Lambda$CDM expansion history.
The contours are the $68$\% and $95$\% CL.
Below we refer to {\it base} as the dataset combination high-$l$ $+$ WP $+$ low-$z$ BAO.
The green contours are found using base $+$ CMASS, the grey contours are found using 
base $+$ WiggleZ, the red contours are founds using  base $+$ 6dFGS, and the 
blue contours show the combined fit to all the growth rate measurements plus the base measurements.
Moreover, we include the AP and BAO information with the growth rate constraints, without
double counting BAO measurements.}
\label{MCMC_gamma_fixed}
\end{figure*}

The first set of results, which assume a $\Lambda$CDM expansion history are 
shown in Fig. \ref{MCMC_gamma_fixed}. This plot shows
the $68\%$ and $95\%$ $2$D likelihood contours for the 
parameter combinations $\{\Omega_{\rm m},\gamma\}$ and $\{\tau,\gamma\}$. 
The expected value of $\gamma$ from GR is given by the grey-dashed line. 
In addition to the growth rate and AP constraints, these measurements 
are inferred using high-$l$ $+$ WP $+$ low-$z$ BAO (which we label in this section as {\it base}).
For fits in this section we do not use the low-$l$ CMB T-T data or
CMB lensing, since we have not included the dependence of these
signals on $\gamma$.
For the final constraint we measure $\gamma =  0.665 \pm 0.0669$, which 
is consistent with GR at the 95$\%$ C.L.

\begin{figure*}
\centering
\includegraphics[width=15.5cm]{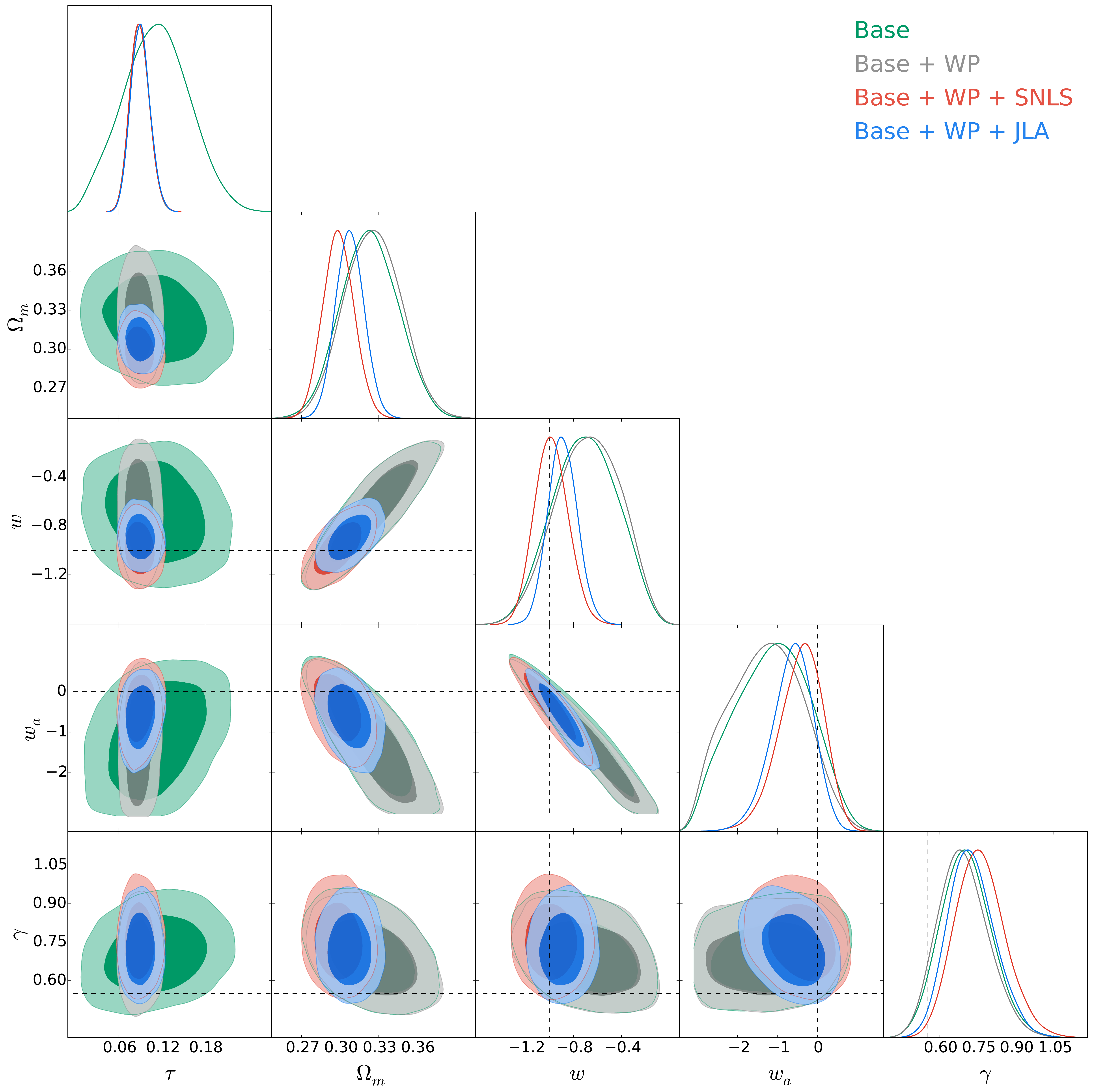}
\caption{$68$\% and $95$\% confidence regions for the most relevant parameters describing model {\color{red}\rom{2}}. 
The base sample of datasets, as refereed to above, represents the combination High-$l$ $+$ $H_{0}$ $+$ RSD/AP $+$ low-$z$ BAO.}
\label{MCMC_gamma_PLOT}
\end{figure*}

The results for ${\bf p_2}$ are presented in Fig.~\ref{MCMC_gamma_PLOT}, where 
we plot the $2$D likelihood contours ($68\%$ and $95\%$), and 
the marginalised $1$D probability distributions for $\gamma,w_{0},w_{\rm a},\Omega_{\rm m},\tau$.
Again, the black-dashed lines indicate the values expected from the 
standard model; namely, $\gamma = 0.55$, $w_{0} = -1$, and $w_{a} = 0$. The degraded 
constraint on $\gamma$ is a direct result of the degeneracy between the expansion and growth 
histories: this is the reason we consider both a fixed and non-fixed expansion history.

We use four different dataset combinations to constrain these parameters, they are defined as follows:
fit 1 is the base sample, fit 2 is base $+$ WP, 
fit 3 is base $+$ WP $+$ SNLS, and fit 4 is base $+$ WP $+$ JLA. We define base here as the combination High-$l$ $+$ $H_{0}$ $+$ RSD/AP $+$ low-$z$ BAO.
We use two SN samples in order to understand how sensitive the growth index is to our choice of adopted dataset.

We will first discuss the main results, which are found using fit 3 and 4 (the red and blue contours in Fig. \ref{MCMC_gamma_PLOT})
and then consider how the constraints are influenced by the different probes.
Using fit 3 we infer (in terms of $68\%$ CL) the marginalised constraints
\begin{eqnarray}
w_{0} &=& -0.98^{+0.13}_{-0.15}  \,, \\
w_{a} &=& -0.42^{+0.62}_{-0.47} \, ,
\end{eqnarray}
which are consistent with the standard model. 
In terms of deviation to the growth history we measure (in terms of $68\%$ CL)
\begin{eqnarray}
\gamma = 0.76^{+0.089}_{-0.087} \,.
\end{eqnarray}
This result is at tension with GR at a level greater than $ 2 \sigma$. Changing our 
SN sample to the JLA sample we find this tension is slightly reduced. Using fit 4
we now measure
\begin{eqnarray}
\gamma = 0.73^{+0.08}_{-0.10} \,,
\end{eqnarray}
which is just consistent at the $2 \sigma$ level, and for the expansion history we 
find $w_{0} = -0.89^{+0.12}_{-0.12}$ and $w_{a} = -0.63^{+0.56}_{-0.45}$.
Note, without including any SN data, using fit 2 (the grey contour in Fig. \ref{MCMC_gamma_PLOT}) 
we measure $\gamma = 0.69^{+0.09}_{-0.11}$,
which is consistent at the $95\%$ C.L. This may suggest there exists a mild tension between 
the growth rate and the SN measurements. Finally, we note 
our measurements of the growth index are relatively insensitive to 
the polarization data, as can be observed in Figure \ref{MCMC_gamma_PLOT} by comparing the 
green (no WP) and grey (including WP) contours.

Comparing the best-fit values for the expansion history using only BAO measurements
with the BAO + SN fit (which is driven by SN) is interesting as it provides a test 
of the significance of non-linear structure on SNe distance measurements \citep{2012MNRAS.426.1121C}.
With fit 2, which only uses the low-redshift BAO measurements to constrain the expansion history,
we infer (in terms of $68\%$ C.L) $w_{0} = -0.68^{+0.29}_{-0.26} $ and $w_{a} = -1.27^{+0.92}_{-0.97}$.
These measurements are consistent at the 95\% CL with the standard model 
and the constraints from the SN + BAO fit; 
moreover, they highlight the current necessity of type Ia SN in placing tight 
constraints on the redshift evolution of the equation of state.  
By introducing the Lyman-$\alpha$ BAO measurements 
into this fit we measure $w_{0} = -0.58^{+0.27}_{-0.22}$ and $w_{a} = -1.55^{+0.74}_{-0.89}$, which
indicates a tension with the standard model predictions at a level $>2\sigma$, in agreement with
the results by \citet{2014JCAP...05..027F}.  Further checks for systematics will be required to confirm this result 
given its significance and the complexity of the measurement.



\subsection{Comparison with Previous Results}

Below we summarise a subsample of previous measurements of the parameters $\{w_{0},w_{a},\gamma \}$. 
\begin{itemize}
\item{
\cite{2014MNRAS.443.1065B} measure $\gamma = 0.772^{+0.124}_{-0.097}$ using
the power spectrum multipoles from the DR11 CMASS sample and Planck: this fit includes the AP effect, but does not 
allow for deviation in the expansion history. This value is consistent with the measurement by \cite{2013MNRAS.433.1202S}
of $\gamma = 0.64 \pm 0.26$ found using the clustering wedges of CMASS combined with BAO and SNe measurements.
}
\item{
\cite{2013MNRAS.432..973R} perform fits to $\{w_{0},\gamma \}$ and $\gamma$. 
For a fixed expansion history, using WMAP combined with galaxy cluster data from {\it ROSAT} and {\it Chandra}, 
they measure $\gamma = 0.415^{+0.128}_{-0.126}$. When adding further data from RSD measurements (WiggleZ and 6dFGS) they find 
$\gamma = 0.570^{+0.064}_{-0.063}$. 
}
\item{
\cite{Beutler:2012fk} measure $\gamma = 0.547 \pm 0.088$ using WMAP$7$ and the two-point correlation function measured from $6$dFGS. 
For this fit the expansion history is fixed, as the AP effect is not relevant. Note, 
there is a small difference between our measurement of $\gamma$ from 6dFGS
and this result. This change is driven by the preference for a higher $\Omega_{\rm m}$ in Planck compared to WMAP.}
\end{itemize}

For this analysis we extend the range of RSD measurements used to constrain $\gamma$
relative to \cite{2013MNRAS.433.1202S,2014MNRAS.443.1065B} and \cite{2013MNRAS.432..973R}. Moreover, relative to 
\cite{2013MNRAS.432..973R} we also use the updated Planck measurements as opposed to WMAP.
The final accuracy of our measurement of the growth index improves upon \citet{2013MNRAS.433.1202S} and \citet{2014MNRAS.443.1065B},
given the additional measurements we analyze.
Note, our constraint on the growth index 
disagrees with \cite{2013MNRAS.432..973R} as we use different datasets, and
the two measurements have similar accuracy as 
we choose to focus only on growth rate measurements from RSD: 
we do not include additional probes sensitive to the growth rate. This position is motivated by 
recent suggestions that there exists some tension between the predictions from a Planck 
cosmology and RSD measurements \citep[e.g.][]{2013PhRvL.111p1301M}.

\section{Conclusions And Discussion}\label{sec:conclusions}

In search of departures from the standard cosmological model and clues towards possible extensions, we 
have measured time- and scale-dependent deviations to the gravitational 
field equations of General Relativity. 
We model these deviations using the time and scale-dependent parameters $\{\gm, \gl\}$. These parameters are defined using 
2 bins in time and 2 bins in scale. $\gm$ modifies the gravitational interaction for non-relativistic particles, 
and hence alters structure formation, while $\gl$ acts equivalently for relativistic particles, 
thus affecting how light propagates through the universe.

To measure the eight parameters describing this model, plus the six describing the standard model,
we utilize a range of cosmological probes including
BAOs, Type Ia SNe,
the CMB, CMB lensing, and the cross-correlation of 
the CMB with large-scale structure probes. In addition, we include measurements of the 
power spectrum multipoles from the WiggleZ and CMASS galaxy redshift samples,
and the velocity power spectrum from 6dFGSv. 
Our motivation for adopting a phenomenological model is to provide a set of 
results that can self-consistently be used to test the widest possible range of models. 
To this end, we have focused on only analyzing measurements on scales within the linear regime. 
We summarise our main results as follows:
\begin{itemize}

\item{We perform a new measurement of the power spectrum multipoles of the WiggleZ survey, 
featuring a new calculation of the window function convolution effects and an improved 
determination of the covariance from N-body simulations.}

\item{Modeling deviation from General Relativity in terms of the growth of large-scale structure, we find the following results, given in 
terms of $68\%$ CLs:
$G_{\rm matter}(z<1;k>0.01)= 0.65\pm 0.43  ,\, G_{\rm matter}(z<1;k<0.01)= 1.22^{+0.39}_{-0.34},\,
G_{\rm matter}(z>1;k>0.01)= 0.53\pm 0.32,\, G_{\rm matter}(z>1;k<0.01)= 0.87\pm 0.30$.
These constraints are consistent with GR (i.e., $\gm = 1$) at the $95\%$ confidence level.
We observe a small tension ($>1\sigma$) for the high-wavenumber and high-redshift bin.}

\item{Modeling deviation from General Relativity in terms of light propagation, we derive the following constraints, 
given in terms of $68\%$ CLs: 
$G_{\rm light}(z>1;k>0.01)= 1.057^{+0.053}_{-0.045},\, G_{\rm light}(z<1;k<0.01)= 1.048\pm 0.048 ,\,  
G_{\rm light}(z<1;k>0.01)= 1.153^{+0.080}_{-0.068},\, G_{\rm light}(z>1;k<0.01)= 1.016 \pm 0.026$. These constraints 
are consistent with General Relativity at the $95\%$ confidence level: the significant improvement in constraining power,
relative to $\gm$, is due to the the sensitivity of the ISW effect and CMB lensing to deviations in $\gl$.}

\item{Adopting an alternative model, we introduce deviation in the expansion and growth histories simultaneously by varying
the growth index and two parameters describing a redshift-dependent equation of state.
For this fit we utilize, among other probes, recent growth rate constraints from RSDs, as measured from the WiggleZ, CMASS, and 6dF surveys.
Our final result assuming a $\Lambda$CDM expansion history (in terms of $68\%$ CL) is $\gamma =  0.665 \pm 0.0669$, while 
allowing the expansion history to deviate from $\Lambda$CDM we measure  $\gamma = 0.69^{+0.09}_{-0.11}$. Both these
results are consistent with the standard model; however, introducing SN measurements to this fit (either SNLS or JLA)
we find a $\sim 2 \sigma$ tension with $\Lambda$CDM.}

\end{itemize}

Probes of the velocity field of galaxies have an indispensable role to 
play in addressing questions of the nature of dark energy as they are uniquely 
sensitivity to only temporal perturbations.
The observational datasets we have analyzed are consistent with a vacuum energy interpretation of dark energy;
however, due to the magnitude of current uncertainties any final conclusions drawn from these, and other current,
observations would be premature. 
In future analysis tomographic weak lensing and galaxy-galaxy lensing measurements will be included to
improve our constraints; furthermore, we will begin assessing the viability of specific models using 
the inferred parameter constraints.

\section*{ACKNOWLEDGMENTS}

AJ and JK are supported by the Australian Research Council Centre 
of Excellence for All-Sky Astrophysics (CAASTRO) through project number CE110001020.
CB acknowledges the support of the Australian Research Council through 
the award of a Future Fellowship.
This work was performed on the gSTAR 
national facility at Swinburne University of Technology. 
gSTAR is funded by Swinburne and the Australian 
Government’s Education Investment Fund.

\vspace{-0.3cm}
\bibliographystyle{mn2e}
\bibliography{Mod_gravity}

\end{document}